\documentclass[superscriptaddress,groupedaddress,nofootnoteinbib,12pt]{article}
\pdfoutput=1
\usepackage{color}
\usepackage{graphicx}
\usepackage{dcolumn}
\usepackage{bm}
\usepackage{amssymb}
\usepackage{amsmath}
\usepackage{sectsty}
\usepackage{colortbl}
\usepackage{latexsym}
\usepackage{float}
\usepackage{ifthen}
\usepackage{enumerate}
\usepackage{url}
\usepackage[force]{feynmp-auto}
\usepackage{jcappub}
    \usepackage{picinpar}
    \usepackage{colortbl}
\usepackage{multirow}
	\usepackage{float}
	      \usepackage{setspace}
\usepackage{array}
\usepackage{bm}
\usepackage{amsopn}

\usepackage{booktabs}
\usepackage[table]{xcolor}
\definecolor{lightgray}{gray}{0.9}

\newcommand{\Mp}{M_{\mathrm{Pl}}}

\newcommand{\ie}{{\it i.~e.}}
\newcommand{\grad}{\nabla}
\DeclareMathOperator{\Or}{O}

\def\d{\textrm{d}}
\def\E{\mathcal{E}}
\newcommand{\p}{\partial}

\def\ba{\begin{eqnarray}}
\def\ea{\end{eqnarray}}
\def\beq{\begin{eqnarray}}
\def\eeq{\end{eqnarray}}

\def\mpl{M_{\rm Pl}}

\def\L{\mathcal{L}}
\def\E{\mathcal{E}}
\def\({\left(}
\def\){\right)}
\def\ie{{\it i.e.}}

\def\nn{\nonumber}
\def\mn{_{\mu \nu}}

\def\p{\partial}
\def\mupn{^\mu_{\ \nu}}
\def\<{\langle}
\def\>{\rangle}

\def\D{\mathcal{D}}

\usepackage{titlesec}
\titleformat{\part}{\Large\bfseries}{}{0pt}{Part \thepart\ --\ }

\newcommand{\para}[1]{\par\vspace{2mm}\noindent\textbf{{#1}}.---}

\newcolumntype{Q}{>{$\displaystyle}l<{$}}
\newcolumntype{q}{>{\columncolor[gray]{0.9}$\displaystyle}l<{$}}
\newcolumntype{R}{>{$\displaystyle}r<{$}}
\newcolumntype{S}{>{$\displaystyle}c<{$}}
\newcolumntype{s}{>{\columncolor[gray]{0.9}$\displaystyle}c<{$}}
\newcolumntype{T}{>{\columncolor[gray]{0.9}}c<{}}

\newsavebox{\tableA}
\newsavebox{\tableB}

\newsavebox{\boxplot}
\newsavebox{\boxplota}

\definecolor{dullpurple}{rgb}{0.431,0.188,0.534}
\definecolor{darkgreen}{rgb}{0.133,0.545,0.133}

\begin{document}

\title{
Riding on irrelevant operators
}
	
\author{Claudia de Rham$^{a,b}$\footnote{Emmy Noether Visiting Fellowship} and Raquel H. Ribeiro$^{a,b}$}
\affiliation{$^{a}$CERCA/Department of Physics, Case Western Reserve University, \\
10900 Euclid Ave, Cleveland, OH 44106, U.S.A.}
\affiliation{$^{b}$Perimeter Institute for Theoretical Physics, \\
31 Caroline St N, Waterloo, Ontario, N2L 6B9, Canada}

	\emailAdd{Claudia.deRham@case.edu}
	\emailAdd{RaquelHRibeiro@case.edu}

\abstract{
We investigate the stability of a class of derivative theories known as $P(X)$ and Galileons against corrections generated by quantum effects. We use an exact renormalisation group approach to argue that these theories are stable under quantum corrections at all loops in regions where the kinetic term is large compared to the strong coupling scale.
This is the regime of interest for screening or Vainshtein mechanisms, and in inflationary models that rely on large kinetic terms.
Next, we clarify the role played by the symmetries. While symmetries protect the form of the quantum corrections, theories equipped with more symmetries do not necessarily have a broader range of scales for which they are valid.
We show this by deriving explicitly the regime of validity of the classical solutions for $P(X)$ theories including Dirac--Born--Infeld
(DBI) models, both in generic and for specific background field configurations.
Indeed, we find that despite the existence of an additional symmetry,
the DBI effective field theory has a regime of validity similar to an arbitrary
$P(X)$ theory. We explore the implications of our results for both early and late universe contexts. Conversely, when applied to static and spherical screening mechanisms, we deduce that the regime of validity of typical power-law $P(X)$ theories is much larger than that of DBI.
}	

\maketitle

\section{Introduction}
\label{sec:introduction}
	
The latest decades have witnessed much effort being put into obtaining theoretical predictions from models
which attempt to describe the relevant processes in either
the early or the late universe (or both).
We often argue that an inflationary period of expansion in the early universe allowed the amplification of quantum fluctuations, which later became imprinted in the cosmic microwave background radiation~\cite{Guth:1980zm}. The statistics
of this anisotropic map have become the principal object of interest
in early universe cosmology,
as they might enable the
reconstruction of the parameters of the microphysical Lagrangian---a process usually referred to as `bottom-up approach.'

Since theories attempting to describe the early universe are quantum by nature, 	a natural question to ask is what sort of operators are generated by radiative corrections to the classical theory and if the theory is indeed stable, and hence both natural and predictive.
If the model is described by an Effective Field Theory (EFT), quantum corrections should not introduce important operators which would then offer additional interaction channels and spoil the classical solutions. If that were to happen, the theory could run out of control, since it would have to be augmented by an infinite tower of operators, from an EFT standpoint.	
The recent results of BICEP2~\cite{Ade:2014xna}, which if confirmed would suggest
a detection of primordial gravitational waves and constrain the tensor-to-scalar ratio, also reinstate the relevance of understanding the merger between inflation models and quantum mechanics.
	
These concerns are not exclusive of inflation and also arise in theories
which model the physics of the late Universe. In particular, to address the current accelerated expansion of the universe, one can argue the dark energy sector responsible for this behaviour consists of one or more light scalars.
These are subject to screening mechanisms that rely on
strong self-interactions and interactions with matter
to effectively hide these light degrees of freedom
from the scrutiny of laboratory and solar system experiments~\cite{Jain:2010ka,deRham:2012az}.
 In this paper we will be interested in a specific type of screening called Vainshtein or kinetic Chameleon~\cite{Vainshtein:1972sx} (see Ref.~\cite{Babichev:2013usa} for a recent review).

 Most if not all the theories exhibiting the Vainshtein mechanism are not typical EFTs since they exhibit the wrong sign for analyticity and include superluminalities~\cite{ArkaniHamed:2002sp,Luty:2003vm}.
These properties imply that they cannot enjoy a standard Wilsonian UV completion\footnote{See Refs.~\cite{Dvali:2010jz,Vikman:2012bx,Alberte:2012is} for alternatives to Wilsonian UV completions.} and EFT arguments might not always be appropriate~\cite{Eden:1966,Landau1959181,Martin:1962rt,Eliezer:1989cr}.
 Nevertheless, because of their useful insight,
 standard EFT arguments are sometimes applied to these theories in the literature.
As such, we shall consider them in this paper within the EFT framework.

    Our focus of interest is to understand whether a specific class of derivative scalar field theories is radiatively stable and to establish the regime of validity of their respective classical predictions.\\

For concreteness, we will explore a special type of theories involving only single derivatives of a light field $\phi$, usually referred to as $P(X)$, where $X=-(\p \phi)^2/\Lambda^4$ and $\Lambda$ is the strong coupling scale.
Such models enjoy a global shift symmetry. These types of theories are especially appealing for models of inflation, where they go by the name of $k$-inflation,
and they were first introduced in Refs.~\cite{ArmendarizPicon:1999rj,Garriga:1999vw}.
There inflation is driven by the non-canonical kinetic term of $\phi$. Since
models inspired by string theory typically produce a nontrivial kinetic structure,
this category of models is indeed extremely interesting.
Moreover, one of the key features of these models
is that the tensor-to-scalar ratio can be enhanced~\cite{Mukhanov:2005bu,Vikman:2006hk}.

$P(X)$ models could also be relevant for the late time acceleration of the Universe (see, for instance, $k$-essence models~\cite{ArmendarizPicon:2000dh,ArmendarizPicon:2000ah,Vikman:2007sj}), where the scalar field can be screened via the Vainshtein mechanism~\cite{Babichev:2009ee}. Indeed, in this paper we shall be interested in exploring these multiple phenomenological facets.

Among the entire class of $P(X)$ theories, the Dirac--Born--Infeld (DBI)~\cite{Tseytlin:1999dj,Silverstein:2003hf,Alishahiha:2004eh,Chen:2005ad} model, where the Lagrangian is roughly
   \begin{equation}
   \mathcal{L}_{\textrm{DBI}}\sim -\Lambda^4\, \sqrt{1-X}\, ,
   \label{eq:DBIsymbolic}
   \end{equation}
has taken a lead role owing to its additional non-linearly realised symmetry, whose infinitesimal form is given by~\cite{deRham:2010eu}
   \begin{equation}
 \phi (x)\longrightarrow \phi(x)+
v_{\mu} x^{\mu} +\phi(x) \, v^{\mu} \partial_{\mu} \phi (x)/\Lambda^4\, ,
\label{eq:dbisym}
\end{equation}
with $x^{\mu}$ labelling the 4-dimensional space-time coordinate.
This symmetry is the remnant in four-dimensions
of a fully realised five-dimensional Poincar\'{e} invariance. DBI has been an extremely popular model for inflation giving rise to large non-gaussianities (see, for instance,
Refs.~\cite{Chen:2005fe,Lidsey:2006ia,Huang:2007hh,Tolley:2008na,Langlois:2008wt,Langlois:2008qf}).
The common prescription for DBI is to assume that its EFT can satisfy the criterion of $|X|\sim 1$ provided the acceleration (which should be properly defined) is small. We will revisit this intuition later, and elaborate on its exact interpretation for different background configurations.
 DBI has also been adopted for models of quintessence or `DBI-essence' in Refs.~\cite{Martin:2008xw,Gumjudpai:2009uy,Zhang:2011za}. \\

Another type of higher derivative theories which also have a reorganised EFT dictated by a hierarchy of derivatives of the field are  Galileon theories, which can arise in a certain limit of massive gravity theories (examples include the Dvali--Gabadadze--Porrati (DGP) model \cite{Dvali:2000hr} and massive gravity \cite{deRham:2010gu,deRham:2010ik,deRham:2010kj}).
Galileon theories are invariant under the transformation
\begin{equation}
	\phi(x) \longrightarrow \phi(x) +v_{\mu} x^{\mu} +c\ ,
	\label{eq:galileansym}
	\end{equation}
where $c$ and $v^{\mu}$ are (scalar and vector) constants.
Guided by this symmetry and the requirement of the absence of ghosts,
the derivative structure of the Galileon Lagrangian is of the symbolic form
\cite{Nicolis:2008in,Deffayet:2009mn}
\begin{equation}
\mathcal{L}_{\textrm{Galileons}}\sim \sum_{n=2}^{5}c_n\  \phi\  \E \E (\partial \partial \phi)^{n-1} \eta^{5-n}\ ,
\label{eq:galileons}
\end{equation}
where $\E$ is the antisymmetric Levi--Civita symbol, $\eta$ refers to the flat (Lorentzian) Minkowski space-time metric, and the contraction of indices is implied.
It is a common statement in the literature that theories described by the Lagrangian \eqref{eq:galileons} have a well defined EFT provided $\partial^n \phi /\Lambda^{n+1}\ll 1$, for $n\geq 3$. We shall revisit this criterion in this paper.

Traditionally, the existence of an additional symmetry (like in DBI and in Galileon theories) is associated with the radiative stability of the model. However, as we shall see in this paper, the symmetry on its own
 is not sufficient to render the theory stable. Neither is the symmetry necessarily required to ensure the radiative stability of the theory.
 The role of the symmetry is rather reserved to protect the derivative
 structure of the terms generated by the radiative corrections, which should, in principle, respect the same symmetry the classical action does.

\subsection*{Summary}
Given the significant progress in developing models both of the early and the late universe, we believe it is timely to revisit their fundamental features as EFTs to fully realise the precision era of cosmology we have recently entered. $P(X)$ theories regroup a large class of these models, which are both theoretically and observationally relevant.
The main regime of interest in such theories is when the kinetic term of the field $\phi$ is large, $|X|\lesssim 1$ (for DBI) and potentially even $|X|\gg 1$ in some other $P(X)$ models.
Then the dynamics is mostly driven by the kinetic structure of the field, rather than its
potential.
The main purpose of this paper is to explore the quantum consistency and classical validity of
$P(X)$ models including DBI field theories in their respective regime of interest. Our results will be focused on $P(X)$ theories for simplicity of the discussion, but can also be applied to theories with higher-order derivative interactions, such as Galileons. We will briefly specify our results for this class of theories---see appendix~\ref{app:cubicgalileon} for more details.

Conventionally, a higher level of symmetry in these models has been associated to a better control of the full theory as a whole (\ie, when including quantum corrections.)
DBI has therefore played a  pivotal role amongst $P(X)$ theories, often claimed to be more `natural' or more `radiatively stable' than an arbitrary model within the $P(X)$ class.
In this manuscript we show that while the symmetry does play a crucial role in preserving a given structure in the quantum corrections,
the symmetry by itself does not change the overall magnitude of these corrections. This implies that models
endowed with more symmetries are not {\it necessarily} more `natural,' and
in particular their regime of validity is not {\it necessarily} larger compared to other $P(X)$ theories.\\

The primary results we have established in this paper are the following:
\begin{itemize}

\item {\bf Regime of validity of the classical solution: a perturbative approach---}Thinking
about DBI as a theory in its own right, it is commonly argued that its classical solutions
    are under control even if $|X| \sim 1$
provided some measure corresponding to an acceleration is small. The reason
behind this belief is that the logarithmic and finite contributions
arising from loops of the field itself involve terms of the form $\p^2 \phi$, which are
assumed to be small within the regime of validity of the theory.

In the first part of the manuscript we quantify this regime of validity of arbitrary $P(X)$ models, based on the same criterion as for DBI and simply ask the question of whether or not symmetries play a crucial role in determining this regime of validity.
We follow a  conventional `covariant' perturbative approach  \`a la  Barvinsky \& Vilkovisky to compute the quantum corrections.  \\

For the specific case of DBI, we show that the result is independent of whether or not the formalism preserves the underlying symmetry.
In particular, in a five-dimensional approach which makes the DBI symmetry manifest, we find the same results as in its four-dimensional counterpart. We also show that contrary to the expectations and despite enjoying an additional symmetry, the regime of validity of DBI classical solutions is typically smaller compared to other $P(X)$ models.

\item {\bf Naturalness and Wetterich exact renormalisation group approach---}Next we address the core of the naturalness question  by considering the  Wetterich exact renormalisation group (ERG) equation, which is valid at all loops and  which at lowest order in a derivative expansion for $P(X)$ takes the form,
    \ba
    \frac{\p P_\kappa(X)}{\p \kappa}=\frac{\hbar}{2} \ {\rm Tr}\left[\frac{\p_\kappa \hat R_\kappa}{\hat R_\kappa+Z^{\mu\nu}_\kappa \p_\mu \p_\nu}\right]\,,
    \ea
where $\hat R_\kappa$ is a regularisation operator, $\kappa$ is the infrared regulator and $P_\kappa$ is the modified effective action at $\kappa$ (also known as effective average action).
The complete exact form of this equation is derived in appendix~\ref{app:app1}. In the above, $Z\sim P'(X)$ is related to the effective kinetic metric in these $P(X)$ models.  The exact expression for $Z$ is given in Eq.~\eqref{eq:defZ}. In the regime of interest (large kinetic term) it follows that $|Z|\gg 1$. This procedure differs from the previous one in that it is exact to all loops and $Z$ is not considered to be a fundamental metric to be introduced in the regularisation scheme.

We solve the full ERG equation by performing a derivative expansion (still {\it  non-perturbatively}, that is, valid at all loops). We find that to all orders in derivatives,
the all-loop quantum contributions introduce negligible modifications to the effective action in the large kinetic term regime where $|Z|\gg 1$ (provided derivatives remain under control).

We can understand this result more intuitively by noticing that the path integral for these theories behaves as
\ba
\int \mathcal{D}[\chi]\ e^{-\frac{1}{\hbar }\int \d^4 x Z^{\mu\nu }[\phi]\p_\mu \chi \p_\nu \chi}\sim \int \mathcal{D}[\chi]\ e^{-\frac{1}{\hbar_{\rm eff} }\int \d^4 x\, (\p \chi )^2}\,,
\ea
where $\chi$ is the field perturbation,
so there is an effective reduced Planck constant, $\hbar_{\rm eff}\equiv\hbar/Z$. In the regime where $|Z|\gg 1$, $\hbar_{\rm eff} \to 0$ and  quantum corrections become irrelevant.

We emphasise that this result is shown to all loops and is non-perturbative.  These results are very different from what one would have guessed following a perturbative prescription, or considering \emph{potential} interactions rather than \emph{kinetic} interactions.
While the analysis focused on $P(X)$ models, it is clear that the results hold for any theory exhibiting the Vainshtein mechanism. Indeed, this paper
highlights a very nontrivial implementation of the Vainshtein mechanism at the quantum level. Such implementations were found previously in Ref.~\cite{deRham:2013qqa} for massive gravity~\cite{deRham:2010kj,deRham:2014zqa}, though in a perturbative version.

Our analysis therefore confirms the naturalness of $P(X)$ models deep within the large kinetic term regime where $|Z|\gg 1$. Importantly, our conclusions are again drawn independently of the fact that
the model might enjoy an additional symmetry, which could in principle cloud the
requirements for naturalness properties.
In fact, our work allowed us to highlight the following facts:
\begin{enumerate}
\item While symmetries are crucial in establishing the form of the quantum corrections, they play little role in naturalness arguments for $P(X)$ theories when the strong coupling scale of the theory does not coincide with the cut-off. In particular, symmetries do not enhance their regime of validity. We emphasise that if we follow a procedure for which DBI does not receive large self-corrections of order of the cut-off then, consistently following the same procedure for an arbitrary $P(X)$ model, implies that terms of the form $X^n$ are \emph{not} generated by quantum effects in $P(X)$.
\item Models relying on a large kinetic term can be made natural deep within their `Vainshtein' region where $|Z|\gg 1$. This is an {\it exact} statement and shows the direct implementation of the Vainshtein mechanism within the loops.
\end{enumerate}
\end{itemize}

\para{Outline}This paper is divided into two parts.
Part~\ref{part:validity} discusses the regime of validity of classical solutions
following a perturbative approach, whereas Part~\ref{part:Naturalness} investigates naturalness considerations
fully non-perturbatively in loops.

In \S\ref{sec:ea} we start by defining essential concepts for this paper, namely the cut-off and the strong coupling scales, relevant and irrelevant operators, and discuss the ambiguities in considering power-law divergences. Readers familiar with these concepts
may wish to proceed directly to
 \S\ref{sec:QC}, where we track finite and logarithmic contributions from loops
 following a conservative viewpoint.
 As a by-product of this analysis, we explore the role of symmetries in these contributions. We derive the regime of validity of tree-level calculations by requiring that the previous quantum contributions are small.  We then apply this criterion to DBI during inflation in \S\ref{sec:early}, and recover a criterion consistent with previous results in the literature. We then move in \S\ref{sec:late} to static and spherically symmetric
 background field profiles, appropriate in screening mechanisms, and compare generic $P(X)$ results with those obtained in DBI and Galileon theories.

Part \ref{part:Naturalness} starts with a discussion of
Wilsonian and effective field actions in \S\ref{sec:Wilsonian}. We revisit the standard question of naturalness and address it using an ERG approach valid at all loops in \S\ref{sec:ERG}.  We establish the naturalness of $P(X)$ theories deep within the high kinetic term regime, which is the regime of phenomenological interest.
We draw a comparison between DBI, Galileons and generic $P(X)$ models.

We briefly summarise our findings in \S\ref{sec:discussion}.
The appendices collect further details about our calculations.
They are organised as follows.
Appendix~\ref{app:app1} contains the derivation the Wetterich ERG equation and it plays a pivotal role in part \ref{part:Naturalness}, while appendix~\ref{app:dimcouplings} includes further details on the derivation of the quantum stability in the large kinetic term regime by solving the dimensionless version of the ERG.
The other appendices collect material which is relevant for part \ref{part:validity}.
Appendix \ref{app:4pf} confirms the results of
\S\ref{sec:QC} by explicit computation of Feynman diagrams.
In appendix~\ref{appendix:moreloops} we generalise the one-loop argument of Part~\ref{part:validity} to higher loops, in appendix \ref{app:cubicgalileon} we derive some relevant results for the cubic Galileon and finally in appendix~\ref{appendix:5d} we provide a complementary derivation of quantum effects in DBI  using a symmetry-preserving five-dimensional approach. \\

\para{Conventions}We will mostly assume (for simplicity) that the background scalar field is living in Euclidean space-time. A generalisation to more arbitrary backgrounds is, however, straightforward, and indeed for the inflationary scenario discussed in \S \ref{subsec:dbiinflation} we will relax this assumption and consider a non-flat, though maximally symmetric, space-time. Greek letters are reserved for space-time indices. Partial derivatives are denoted by $\p$, whilst covariant derivatives are represented by $\nabla$. We use units for which the speed of light and the reduced Planck constant, $\hbar$, are set to unity, except when explicitly said otherwise.
The Planck mass is defined by $\Mp\equiv (8\pi G)^{-1/2}$.

\part{Standard EFT perturbative approach}
\label{part:validity}

We start by computing the quantum corrections to a given single-field model by considering loops from the field itself. Consequently, in the first part of this paper, we will not be addressing the questions of how that theory could have been obtained from integrating out heavy fields, or even naturalness questions such as how high-energy physics affect this  low-energy EFT.  This is where power-law divergences may be used as a surrogate for high-energy effects---we leave this to be explored non-perturbatively in Part~\ref{part:Naturalness}.
 For now, however, we focus on the regime of validity of the field theory by itself  for which it is sufficient to follow only loops of the field, and focus on their logarithmic divergencies.

\section{Effective field theory considerations}
\label{sec:ea}

From a standard standpoint,
EFTs provide a low-energy
insight into the full theory without resolving the
high-energy behaviour. This very appealing
feature relies on the existence of a certain decoupling limit,
which separates high from low-energy phenomena.
At low energies we say that operators with scaling $(E/\Lambda)^\alpha$,
for some $\alpha$, are suppressed by the strong coupling scale $\Lambda$, and therefore
dubbed as \emph{irrelevant}, in the action
\begin{equation}
\mathcal{L}_{\textrm{EFT}} \sim \mathcal{L}_{\textrm{low-energy}}+
\sum_{n>4} c_n \, \dfrac{\mathcal{O}_n}{\Lambda^{n}} \Lambda^4\ ,
\end{equation}
where the operator $\mathcal{O}_n$ has dimensions $\textrm{[mass]}^{n}$ with $n>4$.
The other operators included in
$\mathcal{L}_{\textrm{low-energy}}$ which do not
carry such suppression are, on the other hand, \emph{relevant} operators.
This classification relies uniquely on the mass dimension of the operator,
and its usefulness is linked to the existence of a hierarchy between energy scales.

However, irrelevant operators are not necessarily unimportant.
Indeed, in this paper we will assume a slightly
different way of organizing the EFT expansion of operators,
which has been very common in higher derivative theories (see, for example,
Ref.~\cite{Nicolis:2008in,deRham:2010eu}). For background configurations
which are large (compared to $\Lambda$),
a subclass of operators are no longer suppressed by $\Lambda$, that is,
\begin{equation}
\dfrac{\mathcal{O}_n}{\Lambda^{\alpha_n}}\sim \Or (1)\ \textrm{for some} \ n>4\ .
\label{eq:defirrelev}
\end{equation}
Nevertheless, they are still irrelevant operators
from the standard EFT viewpoint.\footnote{Notice that the reverse is sometimes also true where there can be a subclass of what would have traditionally be a relevant operator which is suppressed and is thus
unimportant in the technical sense. This is especially important in $P(X,\phi)$ theories which will be explored in Ref.~\cite{next}.}
We will see in this paper such a family of operators arising,
and to verify their relevance one needs to check they are \emph{not}
redundant operators, in the technical sense of \emph{not}
generating vanishing equations of
motion. Our principal concern will be to identify the relevant
and irrelevant operators which are quantum mechanically induced and
hence correct the classical Lagrangian.

To summarise and to avoid any confusion in this manuscript
an ``irrelevant operator" refers to an operator which has (mass) dimension greater than $4$ in four dimensions. This is an
operator which is suppressed from the traditional EFT interpretation, but not necessarily from the perspective of the re-organised EFT,
based on the hierarchy between derivatives. If an operator is important in the re-organised EFT we refer to it as ``technically  important.''

\subsection{Cut-off versus strong coupling scale}
\label{sec:covssc}
Before we proceed with the computation of the quantum corrections, it is instructive to recapitulate the concept of regime of validity of the classical field theory.
In the literature the difference between the concepts of
cut-off, $\Lambda_c$, and that of strong coupling scale, $\Lambda$, has sometimes appeared blurred, and so we will define them here. We will also need to introduce the notion of regularisation scale, $\Lambda_r$, and infrared regulator, $\kappa$, which are independent from both the cut-off and the strong coupling scale. The only requirement is that $ \Lambda_r, \kappa < \Lambda_c$ and  $ \Lambda \le \Lambda_c$.

By definition the strong-coupling scale of a theory, $\Lambda$,
 is the scale at which the dominant interactions arise
 and it signals the break-down of perturbative {\it tree-level} unitarity. In a standard EFT approach, at this scale the classical solutions are no longer a good description for the physical system at hand, and quantum corrections (\ie, loops) have to be taken into account.

However, the breakdown of perturbative unitarity does not necessarily imply the breakdown of unitarity and hence new physics. The later scale is the cutoff of the theory, the highest scale at which the EFT can be utilised without introducing new heavy physics. The reason the strong coupling scale and the cut-off are not necessarily the same is that the breakdown of perturbative unitarity only indicates the breakdown of perturbation theory. In a theory with a hermitian Hamiltonian, strongly coupled loop effects may restore unitarity postponing the true breakdown of the EFT to a higher scale. \\

 The concept of
 strong coupling scale is thus very distinct from that of cut-off
 which defines the onset of new physics. The practical implications of identifying the scale $\Lambda$ depend on the theory at hand,
 but the following statements are generically true:
\begin{enumerate}
 \setlength{\itemsep}{0pt}
  \setlength{\parskip}{0pt}
  \setlength{\parsep}{0pt}
\item In many cases,  the strong coupling scale, $\Lambda$, coincides with the onset of new physics,
in which case $\Lambda\sim \Lambda_c$.\vskip0.5cm
\item
However, there can also be a  hierarchy between $\Lambda$ and $\Lambda_c$. At the strong coupling scale, $\Lambda$, different  scenarios may occur and we
highlight that in some of them the theory may still
provide a correct description of the physics at that scale $\Lambda$, if $\Lambda \ll \Lambda_c$. In particular:
\begin{enumerate}
 \setlength{\itemsep}{0pt}
  \setlength{\parskip}{0pt}
  \setlength{\parsep}{0pt}
\item In certain cases it is sufficient to include a finite number of loops
to restore a good description of the microphysical processes at that scale (see, for instance, Ref.~\cite{Aydemir:2012nz} for an instructive `self-healing' example).
\item In most cases an infinite number of diagrams
contributing at the scale $\Lambda$ should be taken into account in order to provide a good description of the physical processes at that scale.
However, this does not mean that the theory necessarily loses predictivity at the scale $\Lambda$. It only signifies that, at that energy, accurate estimates can only be obtained by applying some resummation technique.\\
Physical systems where an infinite number of classes of loop diagrams may be resumed to give finite results (and sometimes even close to classical results) are well known and include Bremsstrahlung scattering (vacuum version of the Cherenkov radiation process)~\cite{Migdal:1956tc}. See also Ref.~\cite{Delbourgo:1970au} for an example in a nonlinar chiral theory.
\item Finally, if an infinite number of loop diagrams ought to be included and
if one can prove that there is no possible converging resummation, then
the theory loses predictivity at the scale $\Lambda$, at least from a {\it standard EFT viewpoint}.
\end{enumerate}
\end{enumerate}

Any theory which relies on irrelevant operators to make classical predictions and exhibits a Vainshtein or screening mechanism must lie within the second set of possibilities,
namely $\Lambda \ll \Lambda_c$. In the past decade, there has been a large interest in models where the strong coupling scale, $\Lambda$, gets redressed by a large background field configuration. If this redressing is to make sense, it is crucial to differentiate between $\Lambda$ and $\Lambda_c$. \\

We conclude this small detour by noting that whilst the estimate of
the cut-off energy scale of the theory can be sometimes ambiguous (since it may be difficult to determine the scale at which other fields ought to be included in the action without knowing the details of the UV completion of the theory),
the strong coupling scale is somewhat easier to assess.
It may indeed vary from the usual method in which one identifies
 the
energy scale contributing in the perturbative
expansion of scattering amplitudes in terms of Feynman diagrams.
As we mentioned before, this happens in cases where a strongly self-interacting background implies a redressing of the interactions, which sometimes has the effect of raising the naive strong coupling scale~\cite{Nicolis:2004qq}.
Given these possible ambiguities, our principal goal is to obtain
results which are explicitly independent of the cut-off of the theory, $\Lambda_c$,
which should render them physically trustworthy.

\subsection{Cut-off dependence and the Wilson action}
\label{sec:powerlaws}

Divergencies in loops appear in the form of power-laws and logarithms.
The central reason for why power-law divergences should not necessarily be trusted as an indication of loop corrections from UV physics, is that the effective action, which controls the physically renormalised quantities, is by definition independent of power-law divergences
(see, for example, Ref.~\cite{Burgess:1992gx}). To understand this we briefly review the Wilsonian picture to renormalisation.

Given a field theory for $\phi$ we define the Wilsonian action $S_{\Lambda_r}(\phi)$ by integrating out all modes in the path integral whose momenta are larger than some $\Lambda_r$, which is the regulator scale.
This can be accomplished by splitting the fields into light and heavy modes, and then the Wilsonian action, $S_{\Lambda_r}(\phi)$, only depends on the modes lighter than $\Lambda_r$.
We must perform this computation in Euclidean signature, which we will keep throughout the remaining of this manuscript.

\para{Universal prediction from the logarithmic term}The Wilson action is given by
\ba
e^{-S_{\Lambda_r}(\phi)}  = \int_{k \ge \Lambda_r} \D [\phi] \, e^{-S(\phi)}\ .
\ea
By construction this action is strongly dependent on the chosen regulator scale $\Lambda_r$. In particular, at one-loop we expect contributions to $S_{\Lambda_r}(\phi)$ which are quartic and quadratic in $\Lambda_r$.  This scale may be chosen arbitrarily and
need not be related with the strong coupling scale, $\Lambda$, nor the cutoff, $\Lambda_c$. However, on the basis of the discussion in \S\ref{sec:covssc}, we do require that $\Lambda_r \le \Lambda_c$ so that the integral on the right hand side is meaningful.

We can then define the Wilson action at another arbitrarily chosen scale $\Lambda_r'< \Lambda_r$ via the finite integral
\ba
e^{-S_{\Lambda_r'}(\phi)}  = \int_{\Lambda_r' \le k \le \Lambda_r} \D [\phi] \, e^{-S_{\Lambda_r}(\phi)} \, .\ea
Again by construction $S_{\Lambda_r'}(\phi)$ is independent of the scale $\Lambda_r$ since we may equivalently define it by the integral
\ba
e^{-S_{\Lambda_r'}(\phi)} = \int_{k \ge \Lambda_r'} \D [\phi] \, e^{-S(\phi)} \, ,
\ea
which is manifestly independent of $\Lambda_r$.
This means that in particular the one-loop divergences that arise in $S_{\Lambda_r}(\phi)$ can be written as
\ba
S_{\Lambda_r}(\phi) \sim \int \d^4 x \,\left[ \Lambda_r^4 \, W_4(\phi) + \Lambda_r^2 \, W_2(\phi) +\ln ( \Lambda_r/\mu) \, W_0(\phi)  + W_{\mu, \rm finite}(\phi)\right]\,,
\ea
where we have chosen an arbitrary sliding scale $\mu$ to define the logarithm.
Crucially the power-law divergencies are automatically cancelled by the loop corrections that arise from integrating out modes between $\Lambda_r'$ and $\Lambda_r$:
\ba
S_{\Lambda_r'}(\phi) = S_{\Lambda_r}(\phi) + \Delta \Gamma_{\Lambda_r' < k < \Lambda_r}\,,
\ea
where
\ba
\Delta \Gamma_{\Lambda_r' < k < \Lambda_r} =- \ln  \int_{\Lambda_r' \le k \le \Lambda_r} \D [\phi] \, e^{-S_{\Lambda_r}(\phi)} \, .
\ea
At one-loop this takes the form
\ba
 \Delta \Gamma_{\Lambda_r' < k < \Lambda_r} \sim  \int \d^4 x && \Bigg[ \Lambda_r'^4 \, W'_4  - \Lambda_r^4 \, W_4+ \Lambda_r'^2 \, W'_2- \Lambda_r^2 \, W_2 +\ln ( \Lambda'_r/\Lambda_r) \, W_0   \nn   \\
&&+ W'_{\mu, \rm finite}-W_{\mu, \rm finite} \Bigg]\, ,
\ea
so that we have
\ba
S_{\Lambda_r'}(\phi) \sim \int \d^4 x \, \left[ \Lambda_r'^4 \, W'_4 + \Lambda_r'^2 \, W'_2 +\ln ( \Lambda'_r/\mu) \, W'_0  + W'_{\mu, \rm finite}\right]
\, .
\ea
Now since by definition $\Delta \Gamma_{\Lambda_r' < k < \Lambda_r} $ is independent of the sliding scale $\mu$, we get an analogue of the Callan--Symanzik equation
for  $\Delta \Gamma_{\Lambda_r' < k < \Lambda_r}$, as follows
\ba
\frac{\p}{\p\mu} \Delta \Gamma_{\Lambda_r' < k < \Lambda_r} =0 \, .
\label{eq:CSeq}
\ea
Then we have $\p_\mu \left( W'_{\mu, \rm finite}-W_{\mu, \rm finite}  \right) =0 $,
and similarly the coefficient of the logarithmic divergence at any chosen regulator scale $\Lambda_r$ is universal
\ba
W_0'=W_0\,.
\ea
Thus the only universal prediction we obtain from the cutoff dependence is the logarithmic term which is captured by the sliding RG scale $\mu$.
Indeed, the standard picture which accompanies the significance of the logarithmic divergencies follows
automatically. Starting at some high energy-scale $\Lambda_r$, Eq.~\eqref{eq:CSeq}
uses the logarithmic running divergence to effectively absorb all the high-energy subprocesses which happen between $\Lambda_{r'}$ and $\Lambda_r$ by sliding the
renormalisation scale $\mu$ from $\Lambda_{r'}$ until it arrives at $\Lambda_r$.
Of course this process can be extended iteratively until all relevant soft microphysics
is encoded in logarithms of large ratios of energy scales and the relevant EFT is obtained. When the logarithms themselves become large, which is rather typical in QCD for example, there are a number of well-known prescriptions which can be applied to make the theory results as competitive as the observational precision at hadron colliders~\cite{Collins:1989gx}.

\para{Effective action}The quantity of interest to us is the effective action, $\Gamma$,
which may be defined in terms of the original action as
\ba
e^{-\Gamma(\phi)} =\int\D [\chi] \, e^{-S(\phi+\chi) + \frac{\delta \Gamma(\phi)}{\delta \phi} \chi} \, .
\ea
Assuming $\phi$ is build out of modes with $k<\Lambda_r$, then the support of $\frac{\delta \Gamma(\phi)}{\delta \phi} \chi $ for $\chi$ modes with $k>\Lambda_r$ is vanishingly small, and similarly for these modes we expect $S(\phi+ \chi) \sim S(\chi)$. Then we have
\ba
 \int\D [\chi] \, e^{-S(\phi+\chi) + \frac{\delta \Gamma(\phi)}{\delta \phi} \chi} &=&  \int_{k<\Lambda_r} \D [\chi]  \int_{k \ge \Lambda_r} \D [\chi]  \, e^{-S(\phi+\chi) + \frac{\delta \Gamma(\phi)}{\delta \phi} \chi}  \nn \\
 &\approx & \int_{k<\Lambda_r}  \D [\chi]  \, e^{-S_{\Lambda_r}(\phi+\chi) + \frac{\delta \Gamma(\phi)}{\delta \phi} \chi} \,,
\ea
and so we may define the effective action in terms of the Wilsonian action defined at an arbitrary scale $\Lambda_r$ as
\ba
e^{-\Gamma(\phi)} = \int_{k<\Lambda_r} \D [\chi] \, e^{-S_{\Lambda_r}(\phi+\chi) + \frac{\delta \Gamma(\phi)}{\delta \phi} \chi} \, .
\ea
Again since by definition
\ba
\frac{\p}{\p \Lambda_r} \Gamma(\phi) =0 \,,
\ea
it follows that all the power-law divergences that arise from one-loop calculations automatically cancel against the power-law divergences in the definition of the Wilson action $S_{\Lambda_r}$. For this reason it is consistent to neglect power-law divergences.

On the other hand the logarithmic terms represent a universal correction that is present even in the infrared limit for $S_\kappa $ with $\kappa \rightarrow 0$. This is the reason why in the first part of this work we shall mainly focus on logarithmic divergences and neglect power-law divergences.
As we mentioned before, when asking naturalness questions power-laws are sometimes viewed as indicators of the high-energy behaviour of the theory. For this reason we shall keep them in the second part of this work when addressing naturalness questions---see Part~\ref{part:Naturalness} for more details.

\section{`Standard' covariant perturbative prescription}
\label{sec:QC}

We start by considering the class of $P(X)$ theories, in which the Lagrangian
only depends on the first derivatives of the scalar field $\phi$
through $X=-(\p\phi)^2/\Lambda^4$. We write
\begin{equation}
S_{E}=\int
\textrm{d}^4 x \ \Lambda^4 \ P(X)\, ,
\label{eq:caction}
\end{equation}
with the understanding that $P$ is some dimensionless function
of $X$ and satisfying
\ba
\label{eq:P(X) small X}
P(X) \to  \frac 12 X \quad {\rm as }\quad |X|\to  0\,.
\ea
The Lagrangian enjoys a global  shift invariance
\begin{equation}
\phi(x) \to \phi(x)+ c\ ,
\label{eq:shiftsym}
\end{equation}
where $c$ is some constant. In some particular cases, the action may have an additional global symmetry such as the DBI symmetry \eqref{eq:dbisym} for
the DBI models given by \eqref{eq:DBIsymbolic}. We remain generic for the rest of this section and consider an arbitrary function $P(X)$.

In the presence of a source, $J$, the classical equation of motion for the field $\phi$ is
\begin{equation}
\frac{2}{\Lambda^4}P''(X) \,\partial^{\alpha}\partial^{\beta}\phi \
\partial_\alpha \phi\, \partial_\beta \phi
- P'(X)\, \Box \phi=J\,.
\label{eq:ceom}
\end{equation}

\subsection{Background field method}
\label{sec:fieldexp}
Expanding the action \eqref{eq:caction}
around a background profile\footnote{We emphasise that this background is only
invoked to compute the effective action and does not need to be a physical background. Similarly the source $J$ need not be a physical source.}, $\phi$, up to quadratic order in the fluctuations,
$\chi$, we find
\begin{equation}
\delta  S_{E}=-\frac{1}{2}
\int{\textrm{d}^4 x}
\Big\{
 Z^{\mu\nu}[\phi] \,
\partial_{\mu} \chi \,
\partial_{\nu} \chi
\Big\}\ ,
\label{eq:deltaS}
\end{equation}
where the kinetic operator, $Z^{\mu\nu}[\phi]$,
only depends on  the field $\phi$ through its first derivatives
\begin{equation}
Z^{\mu\nu}[\phi]=2\, P'(X) \delta^{\mu\nu}-\frac{4}{\Lambda^4}P''(X)\, \partial^{\mu}\phi \partial^{\nu}\phi\,.
\label{eq:defZ}
\end{equation}
As a result, $Z[\phi]$ is manifestly invariant under a global shift.
Notice that the boundary terms can be omitted in this process since they do not contribute to the dynamics.
We include in appendix \ref{app:cubicgalileon} the respective formula for the kinetic operator in Galileon theories for completeness.\\

\para{Regions of interest}For models described by the action \eqref{eq:caction}
the phenomenological regime of interest is that in which $|Z|$ may be large,
that is, when the kinetic term comes to dominate.
In the DBI model, this happens when $|X|\to 1$. In other $P(X)$ models this may occur when $|X|\gg1$. In what follows by `large kinetic term regime' we implicitly assume $|Z|\gg 1$ meaning at least one of the (absolute) eigenvalues of $Z$ is large. We sometimes symbolically refer to this regime as the Vainshtein or screening regime, even though strictly speaking no screening mechanism may occur in that regime. \\

Integrating \eqref{eq:deltaS} by parts, we get
\begin{equation}
\delta S_{\textrm{E}}=\frac{1}{2}\int{\textrm{d}^4 x}\
\chi \,
\Big\{\sqrt{g_{\rm eff}}\,
\ g_{\rm eff}^{\mu\nu}[\phi_0]\, \nabla_{\mu} \nabla_{\nu}
\Big\}\,
  \chi \, ,
 \label{eq:Euclid-action}
\end{equation}
where $g^{\mu\nu}_{\textrm{eff}}$
is defined via the relation
\begin{equation}
\sqrt{g_{\textrm{eff}}}\
g^{\mu\nu}_{\textrm{eff}}\equiv
Z^{\mu\nu}
\,,
\label{eq:effmetric}
\end{equation}
and $\nabla_\mu$ represents the covariant derivative with respect to $g_{\rm eff,\mu\nu}$.
It is clear that
$g^{\mu\nu}_{\textrm{eff}}$ plays the role of (the inverse of) an effective kinetic metric,
with corresponding
determinant in Euclidean space-time
given by $g_{\textrm{eff}}$ which enters in the
integration measure in the action \eqref{eq:Euclid-action}.

\subsection{One-loop effective action}
\label{sec:1PI}
We now compute the one-loop quantum effective action, which is the sum of all
the one-particle irreducible graphs.
The one-loop quantum effective action $\Gamma[\phi]=S[\phi]+ \Gamma^{\textrm{1-loop}}[\phi]$ is a functional of the scalar field $\phi$ and given by (in the Euclidean)
\begin{equation}
e^{-\Gamma^{\textrm{1-loop}}[\phi]}=
\int \mathcal{D}\left[\chi\right]\ \textrm{exp}^{-\frac{1}{2}\chi  \left(\frac{\delta^2 S_{\textrm{E}}[\phi]}{\delta \phi^2}\right) \chi}\,.
\label{eq:qea}
\end{equation}

Starting from the Euclidean action \eqref{eq:Euclid-action} we can write
\begin{equation}
\Gamma^{\textrm{1-loop}}[\phi]= \frac{1}{2}\textrm{log}\ \textrm{det}
\Big\{
\sqrt{g_{\rm eff}[\phi]}\ g^{\mu\nu}_{\textrm{eff}}[\phi]\  \grad_{\mu} \grad_{\nu}
\Big\}\ ,
\label{eq:1loop}
\end{equation}
where `$\textrm{det}$' should be understood as a functional determinant, which represents an infinite sum of Feynman loop diagrams, and provides a (covariant) generalisation to the Coleman--Weinberg effective action~\cite{Coleman:1973jx}.
Notice that this expression is exact
as far as its dependence on the background scalar field profile
goes.

This object can be computed using, for example,
a technique based on the heat kernel expansion~\cite{Parker,Avramidi:2001ns}, which organises the UV divergences as powers of
the local curvature built out of the effective metric in Eq.~\eqref{eq:1loop}.
This technique implicitly uses the metric $g_{\rm eff}$ in the definition of the regularisation scale and the results are manifestly covariant in terms of that metric.  This differs significantly from the approach followed in Part~\ref{part:Naturalness} where the metric $g_{\rm eff}$ is not considered to carry any information about the UV physics.

The power-law divergences are captured by the first two so-called
Seeley--DeWitt coefficients, and the associated quantum corrections read~\cite{Barvinsky:1985an,Barvinsky:1990up}
\begin{equation}
\Gamma^{\textrm{1-loop}}_{\textrm{power-law}}\sim \frac{1}{(4\pi)^2}
\int{\textrm{d}^4 x \sqrt{g_{\textrm{eff}}}} \
\bigg\{\Lambda_c^4
+
\Lambda_c^2 \ \frac{R}{6}
\bigg\}\,.
\label{eq:quantumLpowerlaw}
\end{equation}
Notice that regardless of the specific form of $Z^{\mu\nu}$ these power-law divergencies will always be non-zero both for $P(X)$ and Galileon theories.
At one-loop, the logarithmic quantum contributions are simply given by~\cite{Barvinsky:1985an,Barvinsky:1990up}
\begin{equation}
\Gamma^{\textrm{1-loop}}_{\textrm{log}}\sim -\frac{1}{240(4\pi)^2}
\int{\textrm{d}^4 x \sqrt{g_{\textrm{eff}}}} \
\Big\{
R^2+
2R_{\mu\nu}R^{\mu\nu}
\Big\}\ ,
\label{eq:quantumL}
\end{equation}
where here again the curvature operators are built out of the effective
metric.
This result is due to Barvinsky \& Vilkovisky.
It is clear that this action is manifestly invariant under the global shift symmetry \eqref{eq:shiftsym} present in the classical action \eqref{eq:caction}.

\para{Power-law divergences}The power-law divergences in \eqref{eq:quantumLpowerlaw} are similar in spirit to the renormalisation of the cosmological constant and the Planck scale if we were dealing with a gravitational theory. For our $P(X)$ theory, it is clear that the quartic divergences involves operators of the same form  as $X^n$ as the one present in the original $P(X)$.

 Even in DBI, if $\Lambda \ll \Lambda_c$  and these power-law divergences were taken seriously, one could never access the regime of interest of these theories (large kinetic regime) without  quantum corrections becoming large. Despite the existence of a non-renormalisation theorem for Galileons~\cite{Nicolis:2008in}, the situation is no different there. Indeed, the power-law divergent operators can be made arbitrarily close to the Galileon ones. This means that even for Galileons, one cannot enter the regime of interest (\ie, the Vainshtein region) without being dominated by quantum corrections of the power-law type even if one were to identify $\Lambda=\Lambda_c$.

\label{para:DBIpowerlaws}In the case where we identify $\Lambda=\Lambda_c$, the situation is {\it better for DBI in the five-dimensional embedding} as quartic divergences would simply change the original DBI effective action by order one corrections, but keeping the same DBI structure.
However, in that case we would need to identify the strong coupling scale with the five-dimensional Planck scale and bulk loops would not decouple.
This should be studied with care.

As a result, {\it with the potential exception of DBI}, for all these theories  to make sense in this {\it perturbative} approach---be it Galileons or an arbitrary $P(X)$---the power-law divergences must be {\it unrepresentative} of the UV physics. As discussed in \S\ref{sec:powerlaws} this may well be the case for many theories since power-law divergences are not necessarily good indicators (a similar
viewpoint was expressed by Burgess \& London in Ref.~\cite{Burgess:1992gx}).

In Part~\ref{part:validity} of this paper we will therefore take the approach that power-laws cannot be trusted, and focus solely on logarithmic divergences. This is the approach that needs to be followed {\it perturbatively} for Galileons (and DBI unless $\Lambda=\Lambda_c=M_5$, where $M_5$ is the five-dimensional Planck scale), and it is therefore natural to keep the same one for more generic $P(X)$ models.   We emphasise, however, that this approach is only {\it temporary} and the core of the naturalness problem including power-law divergences will be fully investigated in Part~\ref{part:Naturalness}.

\para{Logarithmic divergences}As justified in the previous arguments,
we now turn to the one-loop logarithmic divergences presented in \eqref{eq:quantumL}.
Crucially, all the operators in Eq.~\eqref{eq:quantumL} involve higher derivatives
compared to the ones in~\eqref{eq:caction}, and they cannot be written
as a simple function of $X$ on its own.

This means that provided we only follow the logarithmic divergences and the finite contributions,
tree-level calculations computed with the original action~\eqref{eq:caction} are under control so long as the higher derivative operators generated in \eqref{eq:quantumL} remain small.
The higher derivative operators depend on the {\it background field}, which implies that the regime of validity of the classical (tree-level) results themselves also depend on the background field configuration.\footnote{We emphasise, however, that this statement is very different from claiming that the cut-off of the theory depends on the background configuration, which if true would violate the decoupling between low and high energy physics typically considered in EFTs.}

In appendix \ref{app:4pf}, we carry out
a one-loop calculation in a specific theory within the
$P(X)$ class in which we keep track of the logarithmic
divergencies, where the derivative structure of the answer in Eq.~\eqref{eq:quantumL} can be seen explicitly.
The generalisation of this result to higher loops is performed in appendix~\ref{appendix:moreloops}. We show that the logarithmic divergences and finite contribution from the higher loops involve even more derivatives and are thus under control provided derivatives are small, and in particular that the one--loop contributions are small.

In what follows we use this criterion to derive the (perturbative) regime of validity of the classical theory.

\subsection{Regime of validity of the classical theory}
\label{sec:non-renorm}

Depending on the context, one may either be interested in a regime where $|X|\lesssim 1$, or allow for a regime where $|X|\gg 1$:
    \begin{itemize}
    \item In the first case where $|X|\lesssim 1$, any operator of the form $X^m  \, \p^n X$ with $n\ge1$ can be made unimportant compared to the classical operators which are all of the form $X^m$, regardless of how large $m$ is.
    \item If we allow for $|X|\gg 1$, the situation is more subtle. Requiring that higher derivatives acting on the field are small may not always be sufficient to effectively suppress  an operator of the form $X^m \p^n X$ when $m\gg 1$. In \S\ref{sec:P(X)screening} we shall provide an example where
    $|X|\gg 1$ and yet the quantum corrections from the field itself combine to remain small subject to higher derivatives being small.
     \end{itemize}

We conclude that for any Lagrangian built out of derivative interactions
involving only first derivatives acting on the field at the level of the Lagrangian, the contributions from the logarithmic and finite parts of the quantum corrections are under control and do not spoil the classical solutions of the theory
as long as we are in a regime where higher derivatives are suppressed. In practise this means that the classical solutions are always under control provided the curvature invariants $R[g_{\rm eff}]$ built out of the effective metric $g_{\rm eff}$ satisfy
\begin{equation}
\sqrt{g_{\rm eff}}\ |R^2[g_{\rm eff}]| \ll \Lambda^4 \ P(X) \, .
\label{eq:criterion}
\end{equation}
This criterion should be applied with care. It is equivalent to the statement that the acceleration in DBI ought to be small, as long as the acceleration is computed appropriately. The unambiguous way of parameterizing this acceleration is discussed in appendix~\ref{appendix:5d}.

The effective metric defined in Eq.~\eqref{eq:effmetric} is conformally related to $Z^{\mu\nu}$ computed in \eqref{eq:defZ}, and we can write
\begin{equation}
\sqrt{-g_{\rm eff}}\ R^2[g_{\rm eff}]
\sim
\left[\(\frac{\p g_{\rm eff}}{g_{\rm eff}}\)^2 +\frac{\p^2 g_{\rm eff}}{g_{\rm eff}}\right]^2
\sim
 \left[\(\frac{\p Z}{Z}\)^2 +\frac{\p^2 Z}{Z}\right]^2\,,
\label{eq:Zloop}
\end{equation}
and the criterion\footnote{We emphasise that this is the condition for the validity of classical results, and it only illustrates the
limitations of the use of classical tools to understand the microphysics at the relevant energy scales. By no means should one ever read this criterion as a `bound' on physical parameters of the theory.} for the validity of the classical solution can thus be symbolically written as
\ba
\label{eq:criteriaZ}
\left|\frac{\p Z}{Z}\right|^4\ ,\  \left|\frac{\p^2 Z}{Z}\right|^2\ll \Lambda^4 P(X)\,.
\ea
We derive the corresponding criterion for Galileons in appendix~\ref{app:cubicgalileon}.\\

Focusing on the requirement \eqref{eq:criteriaZ}, since $Z$ goes as the field velocity, $\p Z$ goes {\it symbolically} as the
local field acceleration.
At this level, we stress two points:
\begin{enumerate}
\item To be more precise, the criterion in \eqref{eq:criterion}
involves the eigenvalues of $Z^{\mu\nu}$. On the other hand, \eqref{eq:criteriaZ} implicitly assumes that
$Z^{\mu\nu}$ is conformally flat, $Z^{\mu\nu}\sim Z \delta^{\mu\nu}$.
One can always choose a basis in which $Z$
is diagonal. However, when there is a hierarchy between the eigenvalues of $Z^{\mu\nu}$,
one needs to ensure that \emph{all} the
combinations of ratios between
the different eigenvalues of $Z$ (which appear in the
expressions for the curvature quantities in the one-loop
effective action) are kept small.
\item The previous expressions are very symbolic, and in particular $\p$ designates the partial derivative if we were in cartesian coordinates of Minkowski. In different coordinate choices, however, the connection should be included. As we shall see, this is especially important when looking at configurations in spherical coordinates  with radius $r$,
as we shall discuss explicitly in \S \ref{sec:late}.
\end{enumerate}

Whether the Lagrangian itself is stable against quantum corrections is yet another question which is related to the naturalness of the Lagrangian and will be addressed in part~\ref{part:Naturalness}.
We notice that nowhere in the derivation of our result have we invoked any symmetry and as such these results are certainly independent of any additional symmetries that may or may not be present in a particular model.

While it is true that some symmetries can protect the structure of the Lagrangian,
they have little to do with their magnitude and with protecting the Lagrangian
and its classical equations of motion from large quantum corrections. For example, given the shift symmetry in $P(X)$ theories, the only requirement imposed by the
presence of this symmetry is that the operators generated by quantum corrections in the effective action obey the same symmetry. However, the symmetry itself is unrelated to the scale at which quantum corrections enter (be it from finite contributions or from divergent pieces).

We explain more explicitly in
appendix~\ref{appendix:5d} how the role of the symmetry enters in DBI models. We follow a fully covariant five-dimensional analysis where the symmetry (five-dimensional diffeomorphism invariance) is manifest. Despite this elegant procedure, which explicitly keeps the symmetry manifest, we recover precisely the same regime of validity for the classical solutions as obtained had we perform the four-dimensional estimation and used the criteria \eqref{eq:criterion}  or \eqref{eq:criteriaZ} without invoking the symmetry.
We illustrate the determination of the regime of validity of the EFT
in specific examples of $P(X)$ theories in the ensuing analysis.

\section{Implications for inflation}
\label{sec:early}

To gain more insight on our results we apply them now to specific classes of models under certain assumptions of the background field configuration. In particular,  we can gauge the impact of our results on inflation model building. In this case, the background field profile is statistically homogeneous and isotropic, and evolves in time. It is its quantum fluctuations which become imprinted in the microwave sky and whose statistics are later observed in the temperature maps. Whichever microphysics operated in the early universe, the same quantum fluctuations which are responsible for structure formation and the temperature anisotropies in the CMB, should also be under control to assure predictiveness of the model.

\subsection{DBI}
\label{subsec:dbigeneral}
The DBI model is explored in more details in appendix~\ref{appendix:5d} where we present its five-dimensional embedding.
We expand the DBI Lagrangian
\begin{equation}
S_{\rm DBI}=\int{\textrm{d}^4 x} \
\left\{
-\Lambda^4 \sqrt{1-X}+\Lambda^4
\right\} \,,
\label{eq:dbiaction}
\end{equation}
where again $X=-(\partial\phi)^2/\Lambda^4$.
We
split the field $\phi$
into a time-dependent background $\phi_0(t)$ and small
inhomogeneous quantum fluctuations,
which propagate with speed of sound
\begin{equation}
c_s^2=\frac{Z^{ii}}{Z^{00}}=\dfrac{P_{,X}}{P_{,X}-2P_{,XX}\ \dot{\phi}_0^2}=1-\dfrac{\dot{\phi}_0^2}{\Lambda^4}\ ,
\end{equation}
where $\dot{\phi}_0$ denotes the derivative of the background field
with respect to the physical time.

One of the most attractive features of DBI is that
the speed of sound of the scalar fluctuations can be made arbitrarily smaller than that of the light when $X=\dot{\phi_0}^2/\Lambda^4$ is arbitrarily close to (but smaller than) unity.
In this case the Lorentz boost factor, defined as
 $\gamma=(1-X)^{-1/2}\equiv c_s^{-1}$,
 can become arbitrarily large. As a result this theory is
 falsifiable since its microphysics signature can be
 significantly constrained by CMB data. In particular,
 Planck data limits non-gaussianity signals which restrict
 $\gamma\lesssim 14$ at 95\% CL~\cite{Ade:2013ydc}.
 This surely means that DBI inflation cannot operate in its
 most interesting regime, where $\gamma\rightarrow \infty$.
 Nevertheless, we take on a conservative approach and
 explore this model from purely theoretical grounds.

Another reason why DBI has been extremely appealing is that it
arises in the context of higher-dimensions and to be
more precise in brane scenarios as a generalisation of the
Nambu--Goto action.
As explained in appendix~\ref{appendix:5d},
we can picture a D3-brane moving in an unwarped space
with $\phi_0$ being the position of the brane relative to the
tip of the throat.
The scalar field $\phi_0$ therefore plays the role of the inflaton,
and the DBI action characterises the motion of the brane in a
generically warped throat.

In this construction, the criterion~\eqref{eq:criterion}
signifies that the brane can move in this higher-dimensional
geometry at a very large speed, but the acceleration of both
the scalar fluctuations as well as the brane itself ought to be small.
Specializing to the logarithmic quantum corrections in
the  action \eqref{eq:quantumL} for the Lagrangian
\eqref{eq:dbiaction}, we impose
\begin{equation}
|\mathcal{L}^{\textrm{1-loop}}_{\textrm{log}}| \ll
|\mathcal{L}_{\rm DBI} |\ .
\label{eq:compareclassquant}
\end{equation}
In the regime of small $c_s$ and focusing
on the most relevant operator, this corresponds to
\begin{equation}
\dfrac{|\dot{\phi}_0^7 \,\ddot{\phi}_0^4|}{\Lambda^{22} c_s^{11}}
\ll c_s
\Lambda^4 \quad \Rightarrow \quad \dfrac{|\ddot{\phi}_0|}{\Lambda^3}
\ll
c_s^{3}\, .
\end{equation}
This estimate is also precisely equivalent
to the condition defined in \eqref{eq:critgamma} using a purely five-dimensional picture
(recalling that $\gamma=1/c_s$).
The condition above is also compatible with the statement usually stated in the literature that the `acceleration' should be small; however, here we make this statement much more accurate.

To conclude, and without loss of generality, the classical inflationary background in DBI can be justified on theoretical grounds whilst being under control provided
\begin{equation}
|\dot{\phi}_0|\sim \Lambda^2 \ \ \textrm{and}
\ \ \dfrac{|\ddot{\phi}_0|}{\Lambda^3}
\sim
\dfrac{|\ddot{\phi}_0|}{\dot{\phi}_0\Lambda}
\ll
\gamma^{-3} \sim 10^{-3}\ ,
\label{eq:dbigenericcriterion}
\end{equation}
where we have assumed that $\gamma$ is as large as possible within the Planck constraints of DBI inflation~\cite{Ade:2013ydc}.
This result is comparable to what happens in screening solutions as we shall see in \S\ref{subsec:dbiscreening}.

\subsection{Application to DBI inflation in de Sitter}
\label{subsec:dbiinflation}
So far we have assumed that the background field lives in
flat Minkowski (or rather Euclidean) space-time. However, if we are to apply these
results to an inflationary setup, we need to consider
the generalisation to an arbitrary space-time background.
In particular, we can assume a de Sitter background,
which not only breaks Lorentz invariance
but also the shift and DBI symmetry \eqref{eq:dbisym}.
We expect the breaking of the symmetry to be quantified by
some power of $H/\Lambda$, and we will make this statement more
precise next.

We adapt our previous results
and write
the classical action
\begin{equation}
S_{E}=\int{\textrm{d}^4 x } \ \sqrt{g} \ P(X)\ ,
\end{equation}
where indices are lowered and raised with respect to the
background metric $g_{\mu\nu}$ and its inverse $g^{\mu\nu}$.
This should not be confused with the effective metric defined in
Eq.~\eqref{eq:effmetric}.

Expanding in perturbations as outlined in Eq. \eqref{eq:defZ}
yields the kinetic operator
\begin{equation}
Z^{\mu\nu}=2\, P_{,X} g^{\mu\nu}-4P_{,XX}\, \partial^{\mu}\phi
\partial^{\nu}\phi\ .
\end{equation}
We can proceed as in \S\ref{sec:fieldexp} and define the following effective metric
\begin{equation}
g^{\mu\nu}_{\textrm{eff}}=\sqrt{\dfrac{Z}{g} }\ Z^{\mu\nu}\ .
\end{equation}
In de Sitter, the explicit computation of the
one--loop effective action (again not trusting the power-laws) shows the first non-redundant operator
which is produced by quantum effects is of the form
\begin{equation}
\mathcal{L}^{\textrm{1-loop}}_{\textrm{log}}
\supseteq \dfrac{H^4}{\left(\sqrt{1-X}\right)^3}\,,
\label{eq:symmbreak}
\end{equation}
where $H$ is the Hubble parameter associated with the de Sitter metric.
Following the requirement
\eqref{eq:compareclassquant},
we conclude that the quantum effects are under control provided
\begin{equation}
\dfrac{H}{\Lambda} \ll c_s \lesssim 1 \quad {\rm and } \quad
\dfrac{|\ddot{\phi}_0|}{\dot{\phi}_0\Lambda}
\ll
\gamma^{-3} \sim 10^{-3}\,.
\label{eq:valdbidS}
\end{equation}
Likewise we can quantify the degree of DBI symmetry breaking
introduced by the de Sitter expansion, which can be read off
from Eq.~\eqref{eq:symmbreak} and is of order
$(H/\Lambda)^4$, with the hierarchy between $H$ and $\Lambda$ being
of order $10^{-2}$.

\section{Implications to screening}
\label{sec:late}
Derivative theories such as the Galileon models introduced in
Ref.~\cite{Nicolis:2008in} have also
seen raised interest
as potential actors in the late time history of the universe.
They can also be relevant for IR modifications of GR like DGP \cite{Luty:2003vm} or massive gravity~\cite{deRham:2010gu,deRham:2010ik}.

We start by investigating screening mechanisms
for $P(X)$ theories and use spherical coordinates, writing the
background profile solution as $\phi(r)$.
In what follows we consider a conformal coupling between the field $\phi$ and an external matter source
at the Planck scale of the form  $\phi T/\mpl$, where $T$ is the trace of the
energy-momentum tensor of the fluid associated with the matter field. This coupling manifestly
breaks the shift symmetry \eqref{eq:shiftsym}, though
very mildly since the coupling is also Planck suppressed.

The most general type of Vainshtein screening mechanism with generalised $P(X)$ models was considered in Ref.~\cite{Babichev:2009ee}. In this section our intention is to illustrate this mechanism and its classical validity by studying two examples: a generic $P(X)$ screening and a DBI screening.
We later compare our results to screening from Galileons.
Some earlier work includes Ref.~\cite{Brax:2012jr} who
focused in obtaining screening solutions. In our paper we rather
explore the consistency of screening solutions within the
framework of a controlled EFT.

\subsection{$P(X)$-screening}
\label{sec:P(X)screening}
Quantum fluctuations play an important role in inflationary theories. Likewise, in theories of late time cosmology,
if a screening solution exists which is capable of
efficiently hide away the presence of the scalar field, $\phi$,
then one ought to be sure that the quantum corrections in that model
are also under control. Below we explore simple cases of Vainshtein-screening which belong to the general class of models explored in Ref.~\cite{Babichev:2009ee}.

Suppose the scalar field interacts with a fixed
point source
distribution through a conformal coupling,
with $T=-M \delta^{(3)}(\vec{r})$.
Then we can show that the equations of motion
can be integrated once
with respect to the radial coordinate to give
\begin{equation}
P'\(-\frac{\phi'(r)^2}{\Lambda^4}\)
\phi'(r)
=
\frac{M}{8\pi \mpl r^2} \,.
\end{equation}
Searching for screening solutions involves obtaining an associated
fifth-force which
ought to be much smaller than the Newton gravitational one
at small enough distances,
while maintaining the Newton square law at large distances. Such solutions will only exist for certain
choices of $P(X)$, but the analysis of quantum
corrections is naturally independent of this choice.

First, we assume $P(X)\to + X/2$ for $|X|\ll1$.  This ensures the correct behaviour at large distances. For a screening mechanism to happen  efficiently, $X$ should either be of order unity or dominate at small distances. Assuming that $X$ is allowed to dominate, $|X|\gg 1$, and that in this strongly coupled regime $P(X)\sim -c_N N^{-1} (-X)^N$, with $c_N$ and $N$ positive constants, then we have
\ba
-X\sim\frac{\phi'(r)^2}{\Lambda^4}\sim\(\frac{M}{8\pi\mpl}\frac{1} {c_N \Lambda^2r^2}\)^{2/(2N-1)}\sim \(\frac{r_*}{r}\)^{4/(2N-1)} \ .
\label{eq:XPXscreen}
\ea
We are implicitly assuming that $P(X)$ is such that one can extrapolate between the free behaviour, $\phi'(r)\sim M/\mpl r^2$ at infinity, to this screened behaviour for small $r$ without any classical instability.

The behaviour~\eqref{eq:XPXscreen} is consistent with  the strong coupling  assumption $|X|\gg1$ provided
$N > 1/2$ and
\ba
\label{eq:StrongCouplingP(X)}
r \ll r_*\equiv\frac{1}{\Lambda}\(\frac{M}{8 \pi c_N \mpl}\)^{1/2}\,,
\ea
where $r_*$ is the strong coupling radius (sometimes also
dubbed Vainshtein, or screening radius).

In this strongly coupled regime, and assuming the effect on a
test-particle of a given mass, we can compare
the magnitude of the force
mediated by the field $\phi$, $F_{\rm \phi}$,
with that of the standard Newton's square law,
 $F_{\rm N}$. We find
\ba
\frac{F_{\rm \phi}}{F_{\rm Newton}}\sim\frac{1}{2 c_N} \(\frac{r}{r_*}\)^{\frac{4(N-1)}{2N-1}} \hspace{10pt}{\rm for}\hspace{5pt} r\ll r_* \,.
\ea
We infer that the screening is effective (in the sense that the force is suppressed compared to the Newton's law) provided
\ba
N >1\,.
\ea
The larger the power $N$ the more efficient
the screening is. For large $N$, the screening behaviour asymptotes to $F_{\rm \phi}/F_{\rm Newton}\sim (r/r_*)^2$ which is as strong a screening as in DBI~\cite{Burrage:2014uwa}. However, as we shall see below, unlike DBI, the regime of validity of this classical $P(X)$-screening solution is much larger, making $P(X)$-screening much more appealing in that respect.

\para{Validity of the EFT}Calculating the local curvature
quantities in the
one-loop effective action \eqref{eq:quantumL} and
imposing \begin{equation}
|\mathcal{L}^{\textrm{1-loop}}_{\textrm{log}}| \ll
|\mathcal{L}^{\textrm{classical}}|\ ,
\label{eq:valEFT}
\end{equation}
we determine that,  the regime of validity of this classical screening solution is
\begin{equation}
r\Lambda  \gg \(\frac{r}{r_\star} \)^{ \frac{{N}}{2N-1}} \,,  \hspace{10pt}\text{or equivalently,}\hspace{10pt}
r\gg \frac{1}{\Lambda}\(\Lambda r_*\)^{-\frac{N}{N-1}} .
\label{eq:constraintPXeg}
\end{equation}
Therefore the background can be very large whilst
satisfying \eqref{eq:XPXscreen},
without the theory running out of control if
Eq.~\eqref{eq:constraintPXeg} is verified. This is
similar in spirit to the regime of validity of theories in
which the background field was only evolving in time, as we
explicitly discussed in \S\ref{subsec:dbigeneral}.

For completeness, we next turn to one of the most popular models within the class of $P(X)$ theories and look into its regime of validity.

\subsection{DBI-screening}
\label{subsec:dbiscreening}

Consider a static, spherically symmetric field profile,
$\phi(r)$, which is governed by the DBI action (with the sign flipped so as to allow screening)
 \begin{equation}
S_{\rm DBI}=\int{\textrm{d}^4 x} \
\left\{
\Lambda^4 \sqrt{1+X}-\Lambda^4
\right\} \,,
\label{eq:dbiactionT}
\end{equation}
which is another special case of the models considered in Ref.~\cite{Babichev:2009ee}.

Assuming again the coupling to matter is conformal
and $T=-M \delta^{(3)}(\vec{r})$,
the solution to the equations of motion satisfies~\cite{Dvali:2010jz}
\begin{equation}
\phi'(r)=\dfrac{\Lambda^2}{\sqrt{1+(r/r_\star)^4}}\ ,
\label{eq:screening}
\end{equation}
where the Vainshtein radius is given by
\begin{equation}
\label{eq:r*DBI}
r_{\star}\equiv \frac{1}{\Lambda}\(\frac{M}{4\pi \mpl}\)^{1/2}\, .
\end{equation}
Here again the Vainshtein radius has the same dependence on the
point source mass, $M$, and the strong coupling scale, $\Lambda$,
as the previous $P(X)$ example \eqref{eq:StrongCouplingP(X)},
and different from the one arising in the case of
the cubic Galileon (though the same as in the quartic
and quintic Galileons).

Screening occurs for small enough $r$, that is, when $r<r_{\star}$,
which corresponds to large $\gamma$, since
\begin{equation}
\gamma\equiv \frac{1}{\sqrt{1+X}}= \(\frac{r_\star}{r} \)^2
\sqrt{1+(r/r_\star)^4}\ .
\end{equation}
Since we are still interested in the regime
corresponding to $\gamma\gg 1$ we will be able to compare the constraints arising
from the validity of the classical solution directly with those from \S\ref{subsec:hdim}, which rely on
higher-dimensional arguments.
Indeed, we are now in a position to fully appreciate
the insights offered when embedding DBI in a higher-dimensional space,
which we have addressed in \S\ref{appendix:5d}.

\para{Validity of the EFT}The condition \eqref{eq:criterion} which is equivalent to \eqref{validity} for DBI is key to understanding the regime of validity of the screening mechanism as $K$ is the invariant measure of the acceleration that transforms appropriately under the DBI symmetry.

For a static and spherically symmetric configuration,  $K\mupn$ is given by
\ba
K\mupn={\rm diag}\(0, \gamma^3\, \frac{\phi''(r)}{\Lambda^2}, \gamma \, \frac{\phi'(r)}{r \Lambda^2},\gamma \,\frac{\phi'(r)}{r \Lambda^2}\)\,.
\ea
The classical screening solution is therefore under control provided\footnote{Ref.~\cite{Burrage:2014uwa} has been updated to reflect these results.}
\ba
\gamma^3 |\phi''(r)| \ll \Lambda^3 \hspace{20pt}{\rm and}\hspace{20pt}
\gamma\, \frac{|\phi'(r)|}{r} \ll \Lambda^3\ ,
\label{eq:constraintsdbir}
\ea
which for the screening solution \eqref{eq:screening} above implies
\ba
r \gg \(\frac{r_*^{2}}{\Lambda}\)^{1/3}= \frac{1}{\Lambda}\(\frac{M}{4\pi \mpl}\)^{1/3}\,,
\ea
or equivalently, to compare with Eq.~\eqref{eq:constraintPXeg}
associated with the generic power-law $P(X)$ model,
\ba
\label{eq:DBIscreeningCriterion}
r\gg \frac{1}{\Lambda}\(\Lambda r_*\)^{2/3} \gg \frac{1}{\Lambda}\,.
\ea
The conditions in Eqs.~\eqref{eq:constraintsdbir} are the
static and spherically symmetric
equivalent
of the conditions obtained in Eq.~\eqref{eq:dbigenericcriterion} for a
time-dependent background profile.
Indeed,  \eqref{eq:constraintsdbir} is a particular case of the criterion derived in Eq.~\eqref{eq:critgamma}.

\subsection{Comparison between screenings}
At this point, one might wonder whether some sub-classes of
$P(X)$ theories are more competitive when we study the range
of scales allowed by their classical description in static
and spherically symmetric profiles.

For comparison purposes, we consider only
the region in parameter space in those models which
gives rise to screening mechanisms.
To make the comparison as generic as possible we
might wish to include the cubic
Galileon~\cite{Nicolis:2008in}. The details of the analysis for the cubic Galileon are provided in appendix~\ref{app:cubicgalileon}.
We start by comparing Eqs.~\eqref{eq:StrongCouplingP(X)} and \eqref{eq:r*DBI} with the result in the cubic Galileon~\cite{Nicolis:2004qq}.
In the cubic Galileon
the scaling of the Vainshtein radius
with the point source mass $M$ is~\cite{Nicolis:2004qq}
\ba
\label{eq:r*cgalileon}
r_*\sim \frac{1}{\Lambda}\(\frac{M}{4\pi \mpl}\)^{1/3}\, .
\ea
Applying the criterion \eqref{eq:valEFT} we find the following
regime of validity for the cubic Galileon
\ba
r \Lambda \gg (\Lambda r_\star)^{-3}\ .
\ea
By inspection of Eqs. \eqref{eq:constraintPXeg} and \eqref{eq:DBIscreeningCriterion} we conclude that,
for these backgrounds,
	Galileon theories have a  broader range of scales for
	which their classical screening solution is under
	control\footnote{By which we mean that the classical operators remain
	unaltered by the ones originated from the logarithmic and  finite contributions
	arising in loops.},
	compared to all $P(X)$ models we considered here, including
	DBI.
	
	Among the $P(X)$ models the ones which are of power-law type
	typically have a larger domain of classical validity than DBI,
	if one relies on the criterion \eqref{eq:valEFT} to determine
	the regime of validity of the EFT.
We reiterate that this is true despite the fact that DBI is motivated by a higher-dimensional construction and enjoys
an additional symmetry compared to generic $P(X)$ models.
This comes to show how subtle the role of symmetry is when applied to these types of considerations.
This is an interesting point worth exploring further which could make screening mechanisms exhibited by $P(X)$ theories as compelling, if not more, compared to DBI models if this is a criterion one values.

\part{Naturalness of $P(X)$ theories}	
\label{part:Naturalness}	

So far we have been focusing on logarithmic (and finite) contributions arising from quantum effects in $P(X)$ theories. However, these considerations had little to say about the naturalness of this class of models. Power-law divergences have indeed been discarded so far for reasons explained previously, but they can be indicative of how low-energy subprocesses are affected by high-energy degrees of freedom.

To address the question of naturalness we now proceed with an exact renormalisation procedure called Wetterich's ERG equation.
This procedure differs from the previous one in three ways.
\label{section:diff_approaches}
First, in this part we remain agnostic about the exact role played by different divergences and keep {\it all} the contributions from quantum corrections (the power-laws, the logarithmic divergences and the finite pieces).
Second, the approach in what follows will be fully non-perturbative making it much more insightful than any perturbative analysis. For instance, a perturbative analysis might find a large one--loop correction to the classical action going as $\Lambda_c^4 X^n$ for a given $n>0$. Stopping there would lead us to deduce that the EFT description would break down when $|X|\sim (\Lambda/\Lambda_c)^{4/n} \ll 1$. However, a fully non-perturbative analysis might give a result going as $\Lambda^4 \(1-(1+\Lambda_c^4/\Lambda^4 X^n)^{-1}\)$ making these non-perturbative contributions irrelevant in the regime where $|X|\gg (\Lambda/\Lambda_c)^{4/n}$.

Finally, a last difference with the approach from Part~\ref{part:validity} is that we do not consider the effective metric \eqref{eq:effmetric} as being fundamental. As a result this metric does not enter in the regularisation scheme (unlike what is implicitly assumed in \S\ref{sec:1PI}) and the result is not manifestly covariant with respect to that metric. We believe this procedure is better justified since we would not expect UV physics to have any knowledge about the low-energy effective metric.

\section{Standard naturalness problems in $P(X)$ theories}
\label{sec:Wilsonian}

Traditionally, there have been two ways to view naturalness problems in field theory.

\para{Heavy mass dependence}One way is to track the dependence on the heavy mass threshold corrections that lie from the first mass states beyond the regime of validity of the EFT. This definition is largely insensitive to field redefinitions and respects both linearly and nonlinearly realised symmetries of the system.

The second is to track the cut-off dependence. In the language of the Wilsonian picture, the idea is to assume that if the EFT has a cutoff $\Lambda_c$, then the theory should be naturally defined by $S_{\Lambda_c}(\phi)$ in the notation of the previous section.

If we take this point of view then the trivial mathematical identity that $\Gamma(\phi)$ should be independent of $\Lambda_r$, even when $\Lambda_r= \Lambda_c$, is turned into a `surprising' fine tuning---it appears necessary to significantly tune the $\Lambda_c$ dependence of the form of $S_{\Lambda_c}$ so that the predicted physical quantities determined by $\Gamma(\phi)$ are not strongly dependent on $\Lambda_c$.

\para{Power-law divergences}The second way to phrase the naturalness problem proceeds as follows.
We start with the classical action \eqref{eq:caction} for $P(X)$ theories.
We take $\Lambda_r=\Lambda_c$ and follow the power-law divergences which, at one-loop, include the following operators
\ba
\label{eq:power1loop}
\L^{\rm 1-loop}_{\Lambda_c}\supset \Lambda_c^4\sum_{n\ge1}\alpha_n X^n+ \Lambda_c^2 \sum_{n\ge 2} \beta_n \p^2 X^n\, ,
\ea
where $\alpha_n$ and $\beta_n$ are dimensionless parameters which only depend on $n$.
 One crucial aspect of these divergencies is that the sum does not truncate (\ie, there is no $N$ for which $\alpha_n=0$ or $\beta_n=0$ for $n>N$). We can get a better insight by performing a wave function renormalisation. The kinetic term is of the form $\mathcal{Z} (\p \phi)^2$. In this one-loop perturbative analysis,  $\mathcal{Z}\sim (1+ \alpha_1 (\Lambda_c/\Lambda)^4)$. We perform a wave function renormalisation by introducing the renormalised field $\phi_R$ defined as
 \ba
 \phi_R\equiv\sqrt{\mathcal{Z}} \phi\,,
 \ea
  and the one-loop contributions go as
  \ba
  \label{eq:power1loop_2}
\L^{\rm 1-loop}_{\Lambda_c}\supset \sum_{n\ge1}\tilde \alpha_n \frac{1}{\Lambda_c^{4(n-1)}}(\p \phi_R)^{2n}+\sum_{n\ge 2} \tilde \beta_n \frac{1}{\Lambda_c^{4n-2}}\p^2 (\p \phi_R)^{2n}\, .
  \ea
 In the large kinetic region, this is worrisome for several reasons. First, the strong coupling scale flows towards the cutoff. Indeed the only relevant scale in \eqref{eq:power1loop_2} is the cut-off, and the original strong coupling scale $\Lambda$ does not even enter.   At higher loops the situation is even worse where the renormalised interaction scale goes as $(\Lambda_c/ \Lambda)^\ell \Lambda\ge \Lambda_c$, where $\ell$ is the number of loops.
  This is often incorrectly used as an argument that the theory cannot be made sense of above $\Lambda$ so that we must take $\Lambda_c \sim \Lambda$. Second, even if we take $\Lambda_c \sim \Lambda$ all powers of $X^n$ receive an order unity modification at the strong coupling scale $\Lambda$ and the functional form of $P(X)$ effectively becomes arbitrary.

As a consequence we would inevitably return to the standard EFT picture that these theories are at best EFTs defined with a cutoff of $\Lambda_c \sim \Lambda$. Even resorting to a symmetry (like in DBI) would not prevent renormalizing the overall coefficient of $P(X)$ to an amount proportional to $\Lambda_c^4$ and again we would need $\Lambda_c\sim \Lambda$ to make sense of that theory.
{\it In the absence of some symmetry} protecting the form of $P(X)$,
the functional form of the $P(X)$ Lagrangian appears uncontrolled.
These perturbative considerations therefore suggest that we cannot trust the classical background as soon as  $|X|\sim 1$.

In the next sections we will argue that even within the cut-off framework, this perspective is too pessimistic, and is an artefact of perturbative arguments. On the contrary, it appears that the large kinetic term region $|Z| \gg  1$ (where $Z^{\mu\nu}$ is defined in Eq. \eqref{eq:defZ}) is precisely the regime where \emph{all} quantum effects are most suppressed whether or not a symmetry is present.

\section{Wilsonian exact renormalisation group}
  \label{sec:ERG}

Up to now, we have seen
that if we work within the Wilsonian picture, and track power-law divergences, then by taking $\Lambda_r>\Lambda$ the loop expansion becomes uncontrolled. This is frequently used to argue that the strong coupling scale, $\Lambda$, must also be the cutoff of the EFT. In reality all this identifies is that perturbation theory which generates the contributions to the loops coming from $k > \Lambda$ is not converging. It may, nevertheless, be possible to find a non-perturbative
method that reorganises the expansion and makes this problem
disappear.\footnote{We emphasise that the techniques we have in mind are very different from
those used in cosmological settings to resum logarithmic contributions by dynamical renormalisation group instruments
 \cite{Burgess:2009bs,Dias:2012qy,Dias:2013rla} (see also Ref.~\cite{Seery:2010kh} for a pedagogical review). In that case the resumation procedure takes care of large distance (IR) perturbative divergences which are not related to the questions addressed in this paper.}\\

The ERG is an exact equation that describes how $S_{\Lambda_r}$ must vary with $\Lambda_r$ so that physical quantities such as $\Gamma(\phi)$ are independent of $\Lambda_r$. This is the approach utilised for example in Polchinski's ERG equation~\cite{Polchinski:1983gv},
and it is widely applied in quantum field theory
and statistical physics contexts (see Ref.~\cite{Berges:2000ew} for a review). However, as we have emphasised, this equation keeps track of the unphysical dependence of $S_{\Lambda_r}$ on the arbitrarily defined regularisation scale which must automatically cancel in the construction of $\Gamma(\phi)$. An approach that is more suitable for our purposes was given by Wetterich which uses the effective action as the fundamental quantity~\cite{Wetterich:1992yh}
(see also Tetradis \& Wetterich~\cite{Tetradis:1993ts}). In brief, this approach introduces an infrared cutoff, $\kappa$, into the definition of the effective action. This is appropriate here since we will be interested in theories such as $P(X)$ models for which the shift symmetry renders them massless making the infrared contribution to the loops problematic.

\subsection{Exact renormalisation group equation}
\label{subsec:ergeq}
The modified definition of the effective action which includes the infrared regulator\footnote{It is interesting to point out here that since we are introducing an IR regulator rather than a UV one, we would ultimately send $\kappa \to 0$ which means there should be no issue promoting this prescription to Lorentzian.} $\kappa$, usually called the \emph{effective average action}, is
\ba
\label{ERG1}
e^{-\Gamma_\kappa(\phi)} =\int\D \chi \, e^{-S(\phi+\chi) + \frac{\delta \Gamma_k(\phi)}{\delta \phi} \chi -\int \d^4 x \frac{1}{2}
\chi \hat R_{\kappa} \chi} \, ,
\ea
and the regularisation operator $\hat R_\kappa$ is chosen to have the following properties
\ba
&& \hat R_\kappa(- \Box) \rightarrow \mathcal{Z}_\kappa \kappa^2  \, , \quad \partial \rightarrow 0  \, ,  \\
&& \hat R_\kappa(- \Box) \rightarrow \infty \, , \, \quad \kappa \rightarrow \infty  \,  ,\\
&& \hat R_\kappa(- \Box) \rightarrow 0 \, , \, \quad \kappa \rightarrow 0  \,,
\ea
where $\mathcal{Z}_\kappa$ is the standard wave function renormalisation, not to be confused with $Z^{\mu\nu}_\kappa$.\\

The  choice of IR regulator $\chi \hat R_\kappa \chi$  in Eq.~\eqref{ERG1} (and in Eq.~\eqref{eq:Wkappa}) acts as a mass term which explicitly breaks the shift symmetry. Notice, however, that it merely regulates the field propagator and does not act as a new interaction. As a result,
there is no change in the Feynman rules associated with this new operator. Consequently, no new, symmetry-violating operators can be generated from this IR regulator. As pointed out in Ref.~\cite{Burrage:2010cu} within the context of Galileons, even though a mass term breaks the shift symmetry, it {\it can still be consistently treated as an irrelevant deformation of a shift-invariant Lagrangian}.

\para{Wave Function Renormalisation}In usual presentations of the ERG it is common to
introduce a wavefunction renormalisation $\mathcal{Z}_\kappa$ to account for anomalous dimensions of the field and for the existence of critical points.
Here the entire function $\tilde P_\kappa(\tilde X_\kappa)$ itself is already a highly nontrivial wave function renormalisation and it would not make sense to define the wave function renormalisation as a function of the field itself.  Rather, we define the wave function renormalisation $\mathcal{Z}_\kappa$ by the behaviour of the theory in the small kinetic term regime, where we define
\ba
\label{eq:wavefunction}
P_\kappa(X) \to \frac 12 \mathcal{Z}_\kappa X + \mathcal{O}(X^2)\,.
\ea
In the small kinetic term regime $|Z^{\mu\nu}_\kappa[\phi]|\sim \mathcal{Z}_\kappa$, whereas in the large kinetic term regime, $|Z^{\mu\nu}_\kappa[\phi]| \gg \mathcal{Z}_\kappa$.
In the case of screening, the choice   \eqref{eq:wavefunction} is equivalent to setting the wave function renormalisation based on the behaviour of the field at infinity which is the only meaningful choice.
\\

\para{Example of regularisation operator}For example, we may take the form
\ba
 \hat R_{\kappa}(-\Box) = \frac{(-\mathcal{Z}_\kappa \Box)}{e^{-\Box/\kappa^2}-1} \, .
\ea
The effect of this operator is to give a mass, and hence infrared cutoff, to the low momenta modes, but leave the high momenta modes (compared to $\kappa$) unaffected.

Despite appearances, the effective average action is related to the Wilsonian action $S_{\Lambda_r}$ by a Legendre transformation ~\cite{Morris:1993qb}, and therefore encodes
the same information. The intuitive reason for this is that in $S_{\Lambda_r}$ we include all contributions for modes with $k>\Lambda_r$, but only tree contributions for modes with $k<\Lambda_r$. Similarly for $\Gamma_k$ we include only loops from modes with $k>\kappa$. The condition $\hat R_\kappa(- \Box) \rightarrow \infty$ as $\kappa \rightarrow \infty $ forces the path integral do be dominated by $\chi=0$ with vanishingly small fluctuations implying
\ba
\lim_{\kappa \rightarrow \infty} \Gamma_{\kappa}(\phi) = S(\phi) \, .
\ea
Alternatively, we may modify the definition of $\hat R_\kappa$ so that $\hat R_\kappa(- \Box) \rightarrow \infty$ as $\kappa \rightarrow \Lambda_c$ so that
\ba
\lim_{\kappa \rightarrow \Lambda_c} \Gamma_{\kappa}(\phi)=S_{\Lambda_c}(\phi) \equiv S(\phi)\,,
\ea
where $S_{\Lambda_c}(\phi)$ is the Wilson action at the cutoff scale, $\Lambda_c$. Implicit in this last statement is the idea that the Wilson action defined at the cutoff is
the natural action to define the EFT. However, we can equivalently choose to define the theory at any scale we choose. In particular, in the case of $P(X)$ models, it is more natural to define the theory at the strong coupling scale, $\Lambda$. \\

From the definition of the effective average action we can derive the ERG equation \cite{Wetterich:1992yh}
\ba
\label{eq:ERG}
\frac{\partial \Gamma_{\kappa}}{\partial \kappa} = \frac{1}{2} {\rm Tr} \left[ \partial_{\kappa} \hat R_{\kappa} \frac{1}{\hat R_{\kappa} + \frac{\delta^2 \Gamma_\kappa}{\delta \phi^2}}\right] \, .
\ea
We give the details of its derivation in Appendix~\ref{app:app1}. This is an exact (all loop orders) non-perturbative renormalisation group equation that contains all the information about a given field theory. It automatically satisfies
\ba
\lim_{\rm \kappa \rightarrow 0} \Gamma_{\kappa}(\phi) \rightarrow \Gamma(\phi)\,,
\ea
and is usually solved with the boundary condition
\ba
\Gamma_{\kappa= \Lambda_c}(\phi) =S_{\Lambda_c}(\phi) \, .
\ea

\para{Connection with the one-loop effective action}This ERG equation can be seen simply as a renormalisation group improved version of the one-loop effective action. To see this we note that if we compute \eqref{ERG1} to one-loop we would obtain
\ba
\Gamma_\kappa(\phi) \approx  S(\phi) + \frac{1}{2} \ln {\rm Det} \left[ \hat R_{\kappa} + \frac{\delta^2 S}{\delta \phi^2} \right] = S(\phi) + \frac{1}{2} {\rm Tr} \ln \left[\hat R_{\kappa} + \frac{\delta^2 S}{\delta \phi^2} \right] \, .
\ea
Differentiating with respect to $\kappa$ gives
\ba
\frac{\partial \Gamma_{\kappa}}{\partial \kappa}  \approx \frac{1}{2} {\rm Tr} \left[ \partial_{\kappa} \hat R_{\kappa} \frac{1}{\hat R_{\kappa} + \frac{\delta^2 S}{\delta \phi^2}}\right] \, .
\ea
This would be the one-loop result.
The ERG improvement corresponds to effectively replacing $S$ on the right hand side of this equation with $\Gamma_{\kappa}$ which then gives us back the ERG equation to all loops.

\para{Choice of Regulator}As in any cut-off regularisation scheme, the answer we obtain is not
typically invariant under field redefinitions. In reality there is an infinite number of possible ERG equations we could derive for a given field theory~\cite{Rosten:2010vm}. For this reason we may choose one best suited to the problem at hand. In particular the choice of regulator should respect the symmetries of the low energy EFT.

To see how this works in the case of a $P(X)$ model, let us make the approximation that $\hat R_{\kappa} =  \mathcal{Z}_\kappa (\kappa^2+ \Box)\, \Theta (\Box+\kappa^2)$. This is a common choice in the literature as an optimised regulator for convergence of the approximate solutions of the ERG equation~\cite{Litim:2000ci}.

\para{Derivative Expansion}We now compute the trace at leading order in a derivative expansion assuming that
\ba
\Gamma_{\rm \kappa}(\phi) = \Lambda^4 \int \d^4 x \, P_{\kappa}(\phi) +\text{higher derivative terms}\,.
\ea
The ERG \eqref{eq:ERG} then gives at lowest nontrivial order in the derivative expansion
\ba
\Lambda^4  \frac{\p P_\kappa(X)}{\p \ln \kappa} = \frac{1}{ (2 \pi)^4} \int_{|k|<\kappa} \d^4 k   \left( \frac{ \mathcal{Z}_\kappa \kappa^2}{\mathcal{Z}_\kappa \kappa^2+(Z^{\mu \nu}[\phi]-\mathcal{Z}_\kappa \delta^{\mu\nu}) k_{\mu} k_{\nu}} \right)\,,
\label{eq:ERGPX}
\ea
where $Z^{\mu\nu}[\phi]$ is defined in \eqref{eq:defZ}, and symbolically, $Z\sim P'(X)$. Since $P(X)$ is a function, we see that the ERG is really an infinite number of equations for the full functional dependence of $P(X)$.

\para{Scale Dependence}It is common to remove the overall scale dependence $\kappa$ by defining $\tilde X_\kappa=-(\p \phi)^2/\kappa^4=X \Lambda^4/\kappa^4$, $\Lambda^4 P_{\kappa}(X) = \kappa^4 \tilde P_{\kappa}(\tilde X_\kappa)$,  and $k^{\mu} = \kappa q^{\mu}$ so that the ERG can be put in the dimensionless form
\ba
\frac{\partial \tilde P_\kappa(\tilde X_\kappa)}{\partial \ln \kappa}+ 4 \tilde P_{\kappa}(\tilde X_\kappa)- 4 \tilde P'_{\kappa}(\tilde X_\kappa) \tilde X_\kappa =  \frac{1}{ (2 \pi)^4} \int_{|q|<1} \d^4 q   \left( \frac{\mathcal{Z}_\kappa}{\mathcal{Z}_\kappa+\(\tilde{Z}_{\kappa}^{\mu \nu}-\mathcal{Z}_\kappa \delta^{\mu\nu}\) q_{\mu} q_{\nu}} \right)\,,
\label{eq:dimensionlessERG}
\ea
where
\ba
\label{eq:Zdimensionless}
\tilde{Z}_{\kappa}^{\mu\nu}=2\, \tilde{P}'_{\kappa}(\tilde X) \delta^{\mu\nu}-\frac{4}{\kappa^4}\tilde{P}''_{\kappa}(\tilde X) \partial^{\mu}\phi \partial^{\nu}\phi \, .
\ea

This formalism is common and extremely useful when looking for fixed points of the RG flow. In this work we shall be interested in another question, namely the amplitude of the quantum corrections in different regimes, for which this dimensionless formalism appears to be less convenient.
Moreover, note that even though Eq.~\eqref{eq:dimensionlessERG} is the most common presentation of the ERG equation, it makes the distinction between $\Lambda$ and $\Lambda_c$ less transparent. Given the arguments in part~\ref{part:validity}, this distinction is critical for this class of theories. To make the notation as close as possible with the one presented in part~\ref{part:validity}, we will attempt to solve the ERG equation in the two limiting cases mentioned below, in its dimensionful form. We include a derivation using the dimensionless couplings in appendix~\ref{app:dimcouplings} for completeness.\\

As it stands, the ERG, be it in its form~\eqref{eq:ERGPX} or  \eqref{eq:dimensionlessERG}, is still too difficult to solve explicitly and we need to make some additional approximations to gain traction. There are two obvious regimes of interest:
\begin{itemize}
\item The normal perturbative region, for which $|X| \ll 1$, so that $P(X)$ may be expanded as a polynomial (assuming analyticity at $X=0$ which is guaranteed from our original assumption in Eq. \eqref{eq:P(X) small X});
\item The large kinetic term region, which is our main interest since this contains the new physics we are seeking traces of.
\end{itemize}
We consider these two cases in turn below.

\subsection{RG flow for small kinetic term regime}

As mentioned before, although elegant, the ERG equation is difficult to solve explicitly. As with other non-perturbative systems of equations (such as the Schwinger--Dyson equations), one can truncate the infinite set of equations at some chosen finite order, and solve the resulting finite system of equations exactly. This is not guaranteed to be a good approximation, but it may allow us to capture certain non-perturbative features of the full theory.

If we are only interested in the small kinetic term region, we may expand $P_\kappa(X)$ as a polynomial
\ba
\Lambda^4 P_\kappa(X) = \Lambda^4 \sum_{n=0}^\infty c_n(\kappa) X^n \, ,
\ea
where $c_1(\kappa)$ is the renormalisation of the kinetic term for the scalar field defined previously as $\mathcal{Z}_\kappa=2c_1(\kappa)$. The other coefficients  $c_n$ with $n>2$ are the interaction coefficients.
The idea here is to truncate this expansion at some order $n=N$, and then insert it into the RHS of the ERG equation~\eqref{eq:ERG}. Then we expand the RHS only to order $N$ and neglect the remaining terms. This reduces the ERG equation to a system of $N$ renormalisation group equations which may be solved exactly or numerically to determine the flow.

\para{Instructive toy-model}We illustrate this method with the simplest possible nontrivial example $N=2$. Notice that this case is also studied in a perturbative
language in terms of Feynman diagrams in appendix~\ref{app:4pf}. For this example it is enough to expand the RHS of the ERG equation to second order in $X$,
\ba
\label{eq:Truncation1}
\Lambda^4\frac{\p P_\kappa(X)}{\p \ln \kappa} &=&   \frac{1}{ (2 \pi)^4} \int_{|k|<\kappa} \d^4 k \left[
1-\frac{\mathbb{X}_\kappa^{\mu\nu}k_\mu k_\nu}{2 c_1(\kappa)\kappa^2}
+\(\frac{\mathbb{X}_\kappa^{\mu\nu}k_\mu k_\nu}{2 c_1(\kappa)\kappa^2}\)^2+\cdots
\right]\\
&=&\frac{2\pi^2}{4(2\pi)^4} \kappa^4\left[1-2\frac{c_2(\kappa)}{c_1(\kappa)} X+5\frac{c_2^2(\kappa)}{c_1^2(\kappa)}X^2+\cdots\right]\,,
\label{eq:Truncation2}
\ea
where we have defined $\mathbb{X}_\kappa^{\mu\nu}= Z^{\mu\nu}_\kappa-\mathcal{Z}_\kappa \delta^{\mu\nu}$.
The first term in the square brackets of \eqref{eq:Truncation2} is just the usual renormalisation of the cosmological constant which we ignore (\ie, absorb into $c_0(\kappa)$). The next terms lead to a renormalisation of the coefficients $c_1$ and $c_2$ following the ERG equation
\ba
\Lambda^4\frac{\p c_1(\kappa)}{\p \ln \kappa} &=& -\frac{\kappa^4}{16\pi^2} \frac{c_2(\kappa)}{c_1(\kappa)}\\
\Lambda^4\frac{\p c_2(\kappa)}{\p \ln \kappa} &=& \frac{5 \kappa^4}{32\pi^2} \frac{c^2_2(\kappa)}{c^2_1(\kappa)}\,,
\ea
which are easily solved in terms of their values at $\Lambda_c$ as follows
\ba
\label{eq:c1(kappa)}
c_1(\kappa)&=&c_1(\Lambda_c)\(1+\frac{9 c_2(\Lambda_c)}{128 \pi^2 c^2_1(\Lambda_c)}\frac{\Lambda_c^4-\kappa^4}{\Lambda^4}\)^{2/9}\\
c_2(\kappa)&=&c_2(\Lambda_c)\(1+\frac{9 c_2(\Lambda_c)}{128 \pi^2 c^2_1(\Lambda_c)}\frac{\Lambda_c^4-\kappa^4}{\Lambda^4}\)^{-5/9}\,.
\ea
The renormalised theory is then (ignoring the constant term going as $c_0(\kappa)$),
\ba
\L_\kappa=-c_1(\kappa)\(\p \phi\)^2+\frac{c_2(\kappa)}{\Lambda^4}\(\p \phi\)^4+\cdots\,.
\ea
We now perform the wave function renormalisation, $\phi = \phi_R/\sqrt{\mathcal{Z}_\kappa}$, with $\mathcal{Z}_\kappa=2c_1$ and get
\ba
\L_\kappa=-\frac 12\(\p \phi_R\)^2+\frac{c_2(\kappa)}{4 c_1^2(\kappa)\Lambda^4}\(\p \phi_R\)^4+\cdots\,.
\ea
The renormalised scale at which the interaction $(\p \phi)^4$ arises is therefore
\ba
\Lambda^4_\kappa= \Lambda^4 \frac{4 c_1^2(\kappa)}{c_2(\kappa)}
= \Lambda^4 \frac{4 c_1^2(\Lambda_c)}{c_2(\Lambda_c)} \(1+\frac{9 c_2(\Lambda_c)}{128 \pi^2 c^2_1(\Lambda_c)}\frac{\Lambda_c^4-\kappa^4}{\Lambda^4}\)\,.
\ea
When\footnote{Here and in what follows, we denote by $\Lambda_\kappa$ the strong coupling scale at $\kappa$. In our notation $\Lambda \sim \Lambda_{\kappa=\Lambda_c}$ (up to order one unimportant factors) so $\Lambda_\kappa$ does flow between $\kappa=\Lambda_c$ and $\kappa=0$ even though $\Lambda$ does not.} $\Lambda \ll \Lambda_c$, and starting at $\Lambda_c$ with $c_1(\Lambda_c)\sim c_2(\Lambda_c)\sim1$ we see that $\Lambda_{\kappa \to 0} \sim \Lambda_c$ as was the case in the perturbative one-loop argument presented in \eqref{eq:power1loop_2}. Notice however that this result is exact at all loops, unlike the perturbative argument which would have inferred a different behaviour at higher loops. We have therefore shown that this ERG method is consistent with the one-loop perturbative result in the weak kinetic term region. We obtain a result which is physically entirely consistent: starting at $\kappa= \Lambda_c$ with interactions $X$ which are already small, $|X|\ll 1$, we see that these interactions become even more irrelevant as we run to lower energy scales.

We now turn to the other regime of interest which is the main attraction for this types of theories, namely
when $|X|\lesssim 1$ or even $|X|\gg 1$. Recall that $X$ is defined as $X\equiv -(\p\phi)^2/\Lambda^4$. From the analysis above, the scale $\Lambda_\kappa$ does flow from $\kappa=\Lambda_c$ to $\kappa=0$. However, in what follows, by `large kinetic region' we will only make an assumption on the behaviour of the field at $\kappa=\Lambda_c$. The real assumption behind the `large kinetic region' is that the magnitude of  at least one of the eigenvalues of $Z^{\mu\nu}_{\Lambda_c}$ is large (compared to unity).

\subsection{Quantum stability of large kinetic term regime}
\label{subsec:largeK}

\subsubsection{Leading order in derivatives}

\label{sec:LeadingOrderDerivatives}

It is the large kinetic region which comes in the description of screening mechanisms or inflationary models with large non-gaussianities. For concreteness let us have in mind screening solutions that work via the Vainshtein effect. These mechanisms rely on the fact that when the kinetic term becomes large, the effective coupling of the scalar to matter becomes small. Qualitatively this is the region for which the eigenvalues of $Z^{\mu\nu}$ defined in Eq.~\eqref{eq:defZ} are large in comparison to unity. To be more precise, by `large kinetic term regime', we have in mind the regime where at least one eigenvalue $Z^{\mu\nu}$ at $\kappa=\Lambda_c$ is large,  symbolically $|Z^{\mu\nu}_{\Lambda_c}|\gg 1$.  In this section we perform the analysis keeping the scale dependence explicit. We find this is the most efficient prescription to answer the question of \emph{when} quantum corrections can be small. See Appendix~\ref{app:dimcouplings} for the derivation using the dimensionless couplings introduced in Eq.~\eqref{eq:dimensionlessERG}.

In this region the ERG at leading order in derivatives may be approximated by
\ba
\Lambda^4  \frac{\p P_\kappa(X)}{\p \ln \kappa} &=&   \frac{1}{ (2 \pi)^4} \int_{|k|<\kappa} \d^4 k   \left( \frac{\mathcal{Z}_\kappa\kappa^2}{\mathcal{Z}_\kappa \kappa^2 + \left[Z_\kappa^{\mu \nu}[\phi]-\mathcal{Z}_\kappa\delta^{\mu\nu}\right] k_{\mu} k_{\nu}} \right)\nn\\
 &\approx&
  \frac{1}{ (2 \pi)^4} \int_{|k|<\kappa} \d^4 k   \left( \frac{\kappa^2}{\hat Z_\kappa^{\mu \nu}[\phi] k_{\mu} k_{\nu}} \right) \, .
  \label{eq:Papprox}
\ea
It is justified to neglect the $\mathcal{Z}_\kappa \kappa^2$ in the denominator as we have done  because the integral is already finite in the IR. We define $\hat Z_\kappa ^{\mu\nu}[\phi] \equiv Z_\kappa ^{\mu\nu}[\phi]/\mathcal{Z}_\kappa$.
The second approximation performed in \eqref{eq:Papprox} is justified if we remain in the large kinetic regime $|\hat Z_\kappa ^{\mu\nu}[\phi]|\gg 1$ for all values of $\kappa$. As we shall see,  $|Z^{\mu\nu}_{\Lambda_c}|\gg 1$ implies $|Z^{\mu\nu}_{\kappa}|\gg 1$, so this is  a consistent approximation. We refer to Appendix~\ref{app:dimcouplings} for a more careful analysis where this simplifying approximation is not made.

We recall that we define our $P(X)$ theory at $\Lambda_c$. This means that $\mathcal{Z}_{\Lambda_c}=1$ (which is of course what was set in the previous example), and so $\hat Z^{\mu\nu}_{\Lambda_c}= Z^{\mu\nu}_{\Lambda_c}$.

If $Z^{\mu\nu}$
is conformal, $Z_\kappa^{\mu\nu}=Z_\kappa \delta^{\mu\nu}$, then the integral is easy to perform. We find
\ba
\Lambda^4  \frac{\p P_\kappa(X)}{\p \ln \kappa} \approx \frac{ \kappa^4}{2^4 \pi^2 \hat Z_\kappa} \, .
\ea
In reality $\hat Z^{\mu\nu}$ is always anisotropic, but it is clear that it is the maximum eigenvalue that will dominate in the denominator, and therefore we  approximate the solution as
\ba
\Lambda^4  \frac{\p P_\kappa(X)}{\p \ln \kappa} \approx \frac{\kappa^4 }{2^4 \pi^2 {\rm Max}[\hat Z^{\mu\nu}_{\kappa}]} \, ,
\ea
where
${\rm Max}[\hat Z_\kappa^{\mu\nu}]$ denotes the maximum eigenvalue of $\hat Z_\kappa^{\mu\nu}=Z_\kappa^{\mu\nu}/\mathcal{Z}_\kappa$.

Now we want to solve this equation assuming that the bare theory defined at the scale $\Lambda_c$ is specified by a function $P_{\Lambda_c}(X)$. A priori the running of the function $P_\kappa(X)$ is highly complicated and involves evaluating the following integral
\ba
\Lambda^4 P_0(X) \approx \Lambda^4  P_{\Lambda_c}(X) - \int_{-\infty}^{\ln \Lambda_c} \d \ln \kappa \  \frac{\kappa^4}{2^4 \pi^2 {\rm Max}[\hat Z_\kappa^{\mu\nu}]}   \,.
\ea
However, to get some insight on this expression, we may start by expanding\footnote{We emphasise, however, that even though we are performing a Taylor expansion, already the first order in this expansion includes an infinite number of perturbative terms, so already the first order in the Taylor expansion goes well beyond the standard perturbative approach.} the integrand in a Taylor series about $\kappa=\Lambda_c$. At leading order in this expansion, we obtain the following contribution
\ba
\label{eq:P0}
\Lambda^4 P_0(X)  \approx \Lambda^4  P_{\Lambda_c}(X) -  \frac{\Lambda_c^4}{2^6 \pi^2 {\rm Max}[Z_{\Lambda_c}^{\mu\nu}]}  +\cdots \, ,
\ea
where we have used the fact that $\hat Z_{\Lambda_c}^{\mu\nu}=Z_{\Lambda_c}^{\mu\nu}$. In the case where the leading contribution going as $\Lambda_c^4/ {\rm Max}[Z_{\Lambda_c}^{\mu\nu}] $ is large, the flow from $\kappa=\Lambda_c$ to $\kappa=0$  is large and the next to leading corrections to this expansion are important. However, in the opposite case where the contribution from $\Lambda_c^4/ {\rm Max}[Z_{\Lambda_c}^{\mu\nu}] $ is suppressed, the flow from $\kappa=\Lambda_c$ to $\kappa=0$ is also suppressed and the approximation \eqref{eq:P0} is then justified, see appendix~\ref{app:dimcouplings} for more details.

The key point is that although the leading contribution $\Lambda_c^4/ {\rm Max}[Z_{\Lambda_c}^{\mu\nu}] $ looks like a large quartic divergence, it is Vainshtein suppressed by a factor of $Z$ which becomes larger as we head into the Vainshtein or screening region (or
correspondingly the relevant region when dealing with $k$-inflation or DBI-inflation). This means that deep inside the large kinetic term region, the
 all-orders-in-loop corrections to the leading order in derivative terms in the effective action can be negligible. We conclude that within the screened region, \ie\ when $Z$ is large, the classical theory is protected from large quantum effects by the Vainshtein mechanism itself.

\para{Power-law example}As an illustrative example, suppose we take the theory defined at the scale $\Lambda_c$ to be polynomial of $N$-th order
\ba
P_{\Lambda_c}(X) =  \sum_{n=0}^N c_n X^n \, ,
\ea
where the $c_n$ coefficients are assumed to be of order unity. Note again that we assume that even at the scale $\Lambda_c \gg \Lambda$, the scale
that enters explicitly in the Lagrangian of the $P(X)$ model is set by the strong coupling scale $\Lambda$ and not $\Lambda_c$.
For large kinetic terms, $|X| \gg 1$, we may approximate $P_{\Lambda_c}(X)  \sim  c_N X^N$, and similarly ${\rm Max}[Z_{\Lambda_c}^{\mu\nu}] \sim c_N X^{N-1}$.

Thus the condition that  contributions
to the effective action  at all loops are negligible is
\ba
\Lambda^4 c_N |X|^N \gg \frac{1}{c_N |X|^{N-1}} \Lambda_c^4\ ,
\ea
which for $c_N \sim \Or(1)$ amounts to
\ba
|X| \gg \left( \frac{\Lambda_c}{\Lambda}\right)^{\frac{4}{2N-1}}\,.
\ea
This condition becomes increasingly easier to satisfy as $N$ increases and in the limit $N \rightarrow \infty$ simply becomes $|X| \gg 1$, \ie, which is automatically satisfied in the large kinetic term region.

\subsubsection{Quantum stability at all orders in the derivative expansion}

The previous analysis has shown that if we truncate the ERG to lowest order in the derivative expansion, then $P(X)$ models that have a power-law growth at large $X$ are generically stable under quantum corrections to all orders in loops in the large kinetic term/screening region $|X| \gg 1$. We now extend this argument to all orders in the derivative expansion. To do this we need to establish how to compute the derivative expansion of the ERG equation.

Returning to the exact form of the Wetterich ERG
\ba
\frac{\partial \Gamma_{\kappa}}{\partial \kappa} = \frac{1}{2} {\rm Tr} \left[ \partial_{\kappa} \hat R_{\kappa} \frac{1}{\hat R_{\kappa} + \frac{\delta^2 \Gamma_\kappa}{\delta \phi \delta \phi}}\right] \, .
\ea
We may equivalent rewrite this by introducing a Schwinger parameter $s$ as
\ba
\frac{\partial \Gamma_{\kappa}}{\partial \kappa} = \frac{1}{2}  \int_0^{\infty} \d s \, {\rm Tr} \left[    \exp\left\{-s \left( \hat R_{\kappa} +
\hat A\right)\right\} \, \partial_{\kappa} \hat R_{\kappa}  \right] \, .
\ea
Here both $\hat R_{\kappa}$ and $\hat A\equiv\frac{\delta^2 \Gamma_\kappa}{\delta \phi \delta \phi}$ are differential operators which in a derivative expansion have a quasi-local form
\ba
\hat A(x, \partial) \left[\delta^{(4)}(x-y) \right]= \sum_{n=0}^{\infty} a_n^{\mu_1 \dots \mu_n}(x) \partial_{\mu_1} \dots \partial_{\mu_n} \delta^{(4)}(x-y) \,,
\ea
where coefficient functions $a_n$ are functions of $\phi$ and potentially all orders of derivatives of $\phi$.

To compute the trace we can use the trick that for any differential operator $\hat O(x,\partial)$ then
\ba
{\rm Tr}[\hat O(x,\partial)] = \int \d^4x \int \frac{d^4 k}{(2 \pi)^4}  \, \hat O(x, \partial_{\mu}+ i k^{\mu} )\,,
\ea
where on the RHS the operator is viewed as acting on unity. This relation is easily proven by using a complete set of position and then momentum states to compute the trace.

This gives
\ba
\frac{\partial \Gamma_{\kappa}}{\partial \kappa} = \frac{1}{2} \int \d^4x \int \frac{d^4 k}{(2 \pi)^4} \int_0^{\infty} \d s \,\left[  \exp\left\{-s \left( \hat R_{\kappa}(k^2-\Box-2 ik^{\mu}\partial_\mu ) +\hat A(x, \partial+ik)\)\right\}\,  \partial_{\kappa} R_{\kappa}(k^2)  \right] \, . \nn
\ea
Denoting $\Gamma_{\kappa} = \int \d^4 x \, {\cal L}_\kappa(x)$ then if we are interested in the Lagrangian at the point $x_*$ we can split the operator in the exponent as
\ba
\left( \hat R_{\kappa}(k^2-\Box-2 ik^{\mu}\partial_\mu ) +\hat A(x, \partial+ik)\right) = \left( \hat R_{\kappa}(k^2 ) +\hat A(x_*, ik)\right) + \hat B(x,x_*,\partial+i k) \, ,
\ea
which defines the operator $\hat B$. The idea of this split is that we assume $\partial$ acts only on $x$ and not on the reference point $x_*$. At the end of the calculation we may then take the limit $x \rightarrow x_*$, and by definition $\hat B$ vanishes if we set $\partial =0$ and $x= x_*$. The  derivative expansion corresponds to expanding in powers of the operator $\hat B$. This is very similar in spirit to the point-splitting
regularisation method which serves to regularise the short distance singularities which appear when two given points are taken to coincide~\cite{Birrell:1982ix}.

The corrections to the effective Lagrangian at the point $x_*$ then take the form
\ba
\frac{\partial {\cal L}_\kappa(x_*)}{\partial \kappa} = \lim_{x \rightarrow x_*}\frac{1}{2} \int \frac{\d^4 k}{(2 \pi)^4} \int_0^{\infty} \d s \, e^{-s(\hat R_{\kappa}(k^2 ) +\hat A(x_*, ik) ) } \sum_{n=0}^{\infty} \frac{s^n}{n!}\(\hat B(x,x_*,\partial+i k)\)^n  \partial_{\kappa} R_{\kappa}(k^2)\nn\,.
\ea
We may now perform the integral over $s$, and using a common, crude choice for the regulator $\hat R_{\kappa} =  \mathcal{Z}_\kappa(\kappa^2+ \Box) \Theta (\Box+\kappa^2)$ we obtain
\ba
\frac{\partial {\cal L}_\kappa(x_*)}{\partial \ln \kappa} = \lim_{x \rightarrow x_*}  \int_{|k| < \kappa} \frac{\d^4 k}{(2 \pi)^4} \sum_{n=0}^{\infty} \mathcal{Z}_\kappa\, \kappa^2 \, (\mathcal{Z}_\kappa\kappa^2+\hat A(x_*, ik) )^{-(n+1)}\(\hat B(x,x_*,\partial+i k)\)^n\,.
\ea
Again working with a theory which is at leading order $ {\cal L}_\kappa(x) = P_{\rm \kappa}(X) + \dots$ then at leading order $\hat A(x_*, ik) = Z^{\mu\nu}_\kappa(x_*)k_{\mu} k_{\nu} + \dots$, and assuming we are in the region with $\hat Z \gg 1$ we have
\ba
\label{eq:derExpansion}
\frac{\partial {\cal L}_\kappa(x_*)}{\partial \ln \kappa} \approx  \lim_{x \rightarrow x_*}  \int_{|k| < \kappa} \frac{\d^4 k}{(2 \pi)^4} \sum_{n=0}^{\infty}  \mathcal{Z}_\kappa\, \kappa^2 \,( Z^{\mu\nu}_\kappa(x_*)k_{\mu} k_{\nu} )^{-(n+1)}\(\hat B(x,x_*,\partial+i k)\)^n   \, .
\ea
This form is finally tractable.

The argument for quantum stability now proceeds as before. If we start with the theory defined at the cutoff scale $\Lambda_c$ to be a pure $P_{\Lambda_c}(X)$ model, then at worst $\hat B $ scales as $\hat B \sim Z_{\kappa} \kappa^2$. Thus, quite regardless of the functional dependence of the RHS, the `worst case' estimate for the magnitude of the contributions to $\L_0$ obtained from running down from $\kappa =\Lambda_c$ yields
\ba
\label{eq:kappa=0}
 {\cal L}_{\kappa=0}(x) \approx \Lambda^4P_{\Lambda_c}(X)+ \Lambda_c^4 \sum_{n=0}^{\infty}   \frac{1}{{\rm Max}[Z_{\Lambda_c}] } b_n \, ,
\ea
where the $b_n$ are order unity functions build out of the first and higher derivatives of the field.

\para{Convergence of the derivative expansion}We expect the sum to converge if the derivative expansion is well defined. The exact criterion behind the validity of the derivative expansion in \eqref{eq:derExpansion} is beyond the scope of this study but one can see that \eqref{eq:derExpansion} involves higher and higher orders of $\p Z/Z$. We therefore expect the sum to converge as long as derivatives are small, $\p \ll \Lambda$. For sake of simplicity, we apply here  {\it without further justification} the same criterion \eqref{eq:criterion} or \eqref{eq:criteriaZ} as that derived in Part~\ref{part:validity}, which ensured that the derivatives were small compared to $\Lambda$.

 It is very possible that this estimate is too conservative. Indeed, the coefficients $b_n$ already include contributions from momenta $k$ of order $\Lambda_c$ so it is {\it very likely} that the derivatives could get arbitrarily close to $ \Lambda_c$, in which case we would only need $|\p Z/Z| \ll \Lambda_c$ rather than the much stronger requirements \eqref{eq:criterion} or \eqref{eq:criteriaZ}. As explained at the beginning of \S\ref{section:diff_approaches}, there are several reasons why the conditions obtained here  could potentially be relaxed compared to that found in Part~\ref{part:validity}.

Then assuming the sum converges,  the conditions that the all-loop contributions are negligible modifications to the effective action in the large kinetic term region, $|Z| \gg 1$, is that
\ba
\label{eq:condition}
\Lambda^4P(X) {\rm Max}[Z] \gg \Lambda_c^4
\,.
\ea
We have therefore generalised the result \eqref{eq:P0} to all orders in the derivative expansion. The condition \eqref{eq:condition} is easier and easier to satisfy as one enters deeper within the `Vainshtein' or large kinetic term region.

\subsection{Application to screening}

To illustrate the previous result, let us  revisit the case of static and spherically symmetric screening introduced in \S\ref{sec:late}, under the same conditions of conformal coupling.
Regardless of whether we are dealing with $P(X)$, DBI, or Galileons\footnote{The static and spherically symmetric approximation breaks down {\it classically} for the quartic Galileon so these considerations do not apply in that case. See Refs.~\cite{deRham:2012fg,Berezhiani:2013dca} for more details.}, for  all these screening mechanisms the criterion \eqref{eq:condition} implies
\ba
\label{eq:conditionScreening}
\frac1{\Lambda} (\Lambda r_*)^p  \ll
 r\ll \frac{1}{\Lambda_c} (\Lambda r_*)^{q}\,,
\ea
where the Vainshtein radius was introduced in \eqref{eq:r*DBI} for
$P(X)$ theories, including DBI, and in Eq.~\eqref{eq:r*cgalileon}
for the cubic Galileon.
{Notice that the lower limit is an estimate on when the sum in Eq.~\eqref{eq:kappa=0} is expected to converge, which is the case if the derivative expansion is well-defined.
Assuming that this sum converges, the upper bound arises from the naturalness requirements deep inside the Vainshtein radius. As such, it might be overly conservative, but it is nevertheless suggestive of the limiting length scales for which this theory is well-defined.}

In Eq.~\eqref{eq:conditionScreening} the coefficients $p$ and $q$ are model-dependent if one were to follow the criterion \eqref{eq:criterion} or \eqref{eq:criteriaZ}; in particular, $q=3/2$ for the cubic Galileon whereas $q=1$ for
generic $P(X)$ models.
The exact expressions of the coefficients $p$ were
derived in Eq.~\eqref{eq:constraintPXeg}.
For the power-law $P(X)$ model
then $p<0$, and in
Eq.~\eqref{eq:DBIscreeningCriterion} for DBI we find $p=2/3$. For
the cubic Galileon, $p=-3$.

For concreteness, let us consider for instance  $\Lambda_c\sim $ eV. This is of course well below the Planck scale, but  still much larger than the strong coupling scale $\Lambda$ usually considered during screening. It would be already a major improvement in our understanding if we were able to push the cut-off scale for these types of theories to values as large as $\sim$ eV.  Actually any value which would be larger than the scale of dark energy ($10^{-3}$eV) should already be considered a success.

  Then with  $\Lambda_c\sim $ eV, the quantum contributions at all-loops
introduce negligible modifications to the effective action within the entire solar system (apart from the regions close enough to dense objects such as the Sun and the other planets). This result suggests that the strong coupling scale, $\Lambda$, could be well separated from the cut-off scale, $\Lambda_c$, which is a remarkable feature in these types of theories which `ride on irrelevant operators.'

The fact that the criterion RHS of \eqref{eq:conditionScreening} is the same for DBI as for $P(X)$-screening and that the LHS is actually tighter for DBI than that for a generic $P(X)$ model suggests once more,
that the additional existence of a symmetry has surprisingly little to do with these considerations. We summarise our results in Table~\ref{table:compascreen}.

\begin{table}[ht]
	\heavyrulewidth=.08em
	\lightrulewidth=.05em
	\cmidrulewidth=.03em
	\belowrulesep=.65ex
	\belowbottomsep=0pt
	\aboverulesep=.4ex
	\abovetopsep=0pt
	\cmidrulesep=\doublerulesep
	\cmidrulekern=.5em
	\defaultaddspace=.5em
	\renewcommand{\arraystretch}{1.8}
	\begin{center}
		\small
\rowcolors{1}{}{lightgray}
		\begin{tabular}{ccccc}
			\toprule
			model&\quad
			&
			Lagrangian
			&\quad& regime of validity of the EFT
			\\
			\cmidrule{1-5}
			\cellcolor{lightgray} Galileons & \cellcolor{lightgray} \quad   & \cellcolor{lightgray}
			$-1/2\, (\p\phi)^2+(\p\phi)^2 \, \Box\phi/\Lambda^3$
            & \cellcolor{lightgray}\quad &
			\cellcolor{lightgray}
			$\frac{1}{\Lambda}\(\Lambda r_*\)^{-3}\ll r \ll \frac{1}{\Lambda_c} \(\Lambda r_*\)^{3/2}$
			\\[2mm]
			\cmidrule{1-5}
			 $P(X)$&\quad  &
			 $\sim \left\{\begin{array}{lcc}
            -1/2 (\p \phi)^2 &\quad {\rm for }& X\ll1 \\
            - \Lambda^4 (-X)^{N} &\quad {\rm for }& X\gg1
            \end{array}\right.$
             &\quad
             & $\frac{1}{\Lambda}\(\Lambda r_*\)^{-\frac{N}{N-1}}\ll r \ll \frac{1}{\Lambda_c} \(\Lambda r_*\)$
			\\[2mm]
			\cmidrule{1-5}
			\cellcolor{lightgray} DBI&\cellcolor{lightgray}\quad  &
			\cellcolor{lightgray}
			$\Lambda^4 \sqrt{1+X}-\Lambda^4$
            &\cellcolor{lightgray}\quad & \cellcolor{lightgray}
			$\frac{1}{\Lambda}\(\Lambda r_*\)^{2/3}\ll r \ll \frac{1}{\Lambda_c} \(\Lambda r_*\)$
			\\
 			\bottomrule
		\end{tabular}
	\end{center}
	\caption{Comparison between regimes of validity of different derivative theories (including when the theory is technically natural)
	determined as a function of range of scales.
Note that $r_\star$ scales slightly different
with the mass of the matter distribution which sources the background
field from
model to model as cautioned before. Any screening solution has $\Lambda r_*\gg 1$. In the $P(X)$ model we have $N>1$ (and potentially $N \gg 1$).
 The lower side of the regime is determined by requiring that the derivative expansion converges, using Part~\ref{part:validity} as an indicator. It is likely that the LHS of these criteria are overly restrictive and could be relaxed significantly, as cautioned in the main text.	\label{table:compascreen}}
	\end{table}

\subsection{Background vs. perturbed-field EFT}

So far we have centered our analysis on the question of naturalness. For this we have focused on the EFT of the `background' field $\phi$,
which we have found to be valid both when the kinetic term is small ($|X|\ll 1$)
and when the kinetic term is large and
the criterion \eqref{eq:condition} is satisfied provided the derivative expansion is under control.
It does not mean, however, that the EFT as a whole is valid in all these regimes. The EFT of the background field can be under control and quantum corrections to the background EFT may be small, but this does not yet mean that the perturbed field $\chi$ living on the background determined by $\phi$ is weakly coupled and that quantum corrections are not important to determine its scattering or evolution.

When the EFT for the perturbed field $\chi$ is valid is a separate question which may involve the redressed strong coupling scale as computed for instance in Ref.~\cite{Nicolis:2004qq} for the cubic Galileon. Yet again, as explained in \S\ref{sec:covssc}, the redressed strong coupling scale which determines the breakdown of tree-level unitarity for the perturbations is well distinct from the cut-off.\footnote{As mentioned in \S\ref{sec:covssc}, it could be that the strong coupling scale coincides with the cut-off, but if one is to even talk about `redressed' strong coupling scale, it not only means that the original strong coupling scale has to be well-separated from the cut-off scale but it also means that the strong-coupling scale is independent from the cut-off. Indeed, the cut-off of the theory, \ie, the onset of new physics cannot depend on the background behaviour of the low-energy theory without violating decoupling between low and high energy physics. }
Moreover, the break-down of tree-level unitarity at the (redressed) strong coupling scale does not necessarily mean a loss of predictivity of the theory.

For a power-law $P(X)$ screening of the form $P(X)=X/2 - a_N (-X)^{N}$, we expect the redressed strong coupling scale to go as $\Lambda_* \sim \Lambda X^{N/4}\sim (r_*/r)^{1/2}\Lambda $ in the limit of large $N$.

For DBI, on the other hand, there are some higher order operators which are enhanced by higher powers of the Lorentz factor, and we expect the redressed strong coupling scale to go instead as $\Lambda_* \sim \Lambda / \gamma^{1/4} \sim (r/r_*)^{1/2} \Lambda$ which would make the redressed strong coupling scale \emph{smaller} in screened region. This is an interesting effect due to the square root structure of DBI. In DBI it is therefore particularly important to dissociate the cut-off scale and the (redressed) strong coupling scale.

\section{Summary and discussion}
\label{sec:discussion}

This paper has addressed two essential questions
in a class of derivative Lagrangians, usually known as $P(X)$ models.
These theories are of special interest when the irrelevant operator $X=-(\p\phi)^2$ is large, or at least of order
unity. In this regime we are
`{\it riding on irrelevant operators}' which can be worrisome  from a standard EFT viewpoint. Such operators are important if they are governed by a scale $\Lambda$ which is much smaller than the cutoff of the theory. This immediately begs the question of whether or not the EFT of $P(X)$ models can ever be under control against quantum corrections, meaning whether the renormalised action is close to (or even overrides) the classical action.

We have addressed this question following two different
procedures proposed in the literature:
\begin{enumerate}
\item {\bf Covariant and perturbative approach \`a la Barvinsky \& Vilkovisky---}In
this first part, we
ignored the power-law divergences arising from quantum effects.
We justified this treatment in depth emphasising that it is appropriate if we do not ask a naturalness question from integrating out heavier fields, but are only interested in the quantum corrections from the field itself.
 We find that classical solutions are under control as long as higher derivatives of $X$ are suppressed, or more precisely provided $(\p^2 Z /Z)^2 \ll \Lambda^4 P(X)$. We derived the explicit (covariant) criterion for the suppression of quantum effects and applied it to different contexts:
\begin{itemize}
    \item First, during inflation we recovered the standard result for the regime of validity of DBI inflation amounting to the acceleration of the field being small.
    \item Second, in static and spherically symmetric screening setups. We compared the screening mechanisms for a `generic' power-law $P(X)$ screening to that of DBI, and  have shown that generic $P(X)$ screenings can have a larger regime of validity for their respective classical background solution. The comparison between screenings in different models is summarised on Table~\ref{table:compascreen}.
 \end{itemize}

\item {\bf Exact Wetterich renormalisation group procedure and addressing the naturalness question---}In the second part of this work we have applied an exact {\it all loops} renormalisation procedure and have addressed the core of the naturalness question for generic $P(X)$ models. In this approach we have kept {\it all} the contributions from the quantum corrections, including the power-law and logarithmic divergences, as well as finite pieces.

The ERG approach shows the direct implementation of the `Vainshtein' mechanism in the renormalised effective action. It serves as a suppression mechanism for the quantum effects at all-orders in the loops. We emphasise that this procedure is unrelated to that of the redressed strong coupling scale. Instead, following an ERG approach we find that the new operators in the renormalised effective action are suppressed by a factor of $1/Z$ where $Z\sim P_{,X}$, and $|Z|\gg1$ in the region of interest for this type of theories.

    This proves the full quantum stability of the theory in the regime where the kinetic term is large, $|Z|\gg 1$. $P(X)$ theories are therefore more and more natural as one enters that regime. The same would apply to other theories which exhibit the same type of `large kinetic term regime', like Galileons. Indeed,
    similar conclusions were drawn
    by Brouzakis et al.~\cite{Brouzakis:2013lla,Brouzakis:2014bwa}
    in galileon theories using the heat kernel technique,
    and by Codello et al.~\cite{Codello:2012dx} within a braneworld setup.

    For completeness, we have also considered the less interesting regime, for which $|X|\ll 1$, where the conclusions match that of the perturbative approach at one loop.

\item {\bf The role of symmetries---}In this work we kept a close look at the potential role played by symmetries in these questions of naturalness and `validity of the classical solution.'

We found that the symmetry does of course play a crucial role in repackaging the quantum corrections in a way which preserves the symmetry (this was performed in DBI using a five-dimensional embedding approach). Nevertheless, this nice repackaging of the quantum structure does not say much about the overall order of magnitude of the quantum corrections. As a result when the strong coupling scale does not coincide with the cut-off scale, DBI enjoys the same renormalisation features as any other $P(X)$ theories.
In fact, deep in the high kinetic term region, DBI  is as natural  as any other $P(X)$ model, despite the presence of an additional symmetry.
\end{enumerate}
\vspace*{0,5cm}

To conclude, the net effect of most calculations in derivative Lagrangians has produced a remarkable change in our understanding of the way their EFTs are organised, which relies on the hierarchy between scales being addressed as a derivative hierarchy.
The results in this paper could have profound consequences for these types of theories in general, including Galileon and other models exhibiting the Vainshtein mechanism~\cite{Babichev:2009ee}. See also Refs.~\cite{Creminelli:2013fxa,deRham:2013hsa,deRham:2014lqa,Kampf:2014rka} for related considerations in Galileon theories.

The Vainshtein mechanism relies on non-linear kinetic interactions being important below the cut-off. The principal result of this paper is precisely that the quantum consistency of these theories is tied with these important kinetic interactions. Incorporating the Vainshtein mechanism within the loops themselves has uncovered a mechanism by which quantum corrections are under control. This can open the venue for more models to be taken seriously in model building, both during inflation and late time acceleration.

	\acknowledgments
	We thank Lasha Berezhiani, Justin Khoury, Alberto Nicolis, and especially David Seery and Andrew J.~Tolley for useful feedback and comments on the manuscript. We also especially thank Andrew J.~Tolley for very fruitful discussions on the ERG.  \\
	CdR and RHR are
	supported by a Department of Energy grant DE-SC0009946.
	RHR would like to thank DAMTP (Cambridge, UK)  for hospitality
	and the Perimeter Institute for
	Theoretical Physics (Waterloo, Canada)
	for hospitality and support whilst this work was in progress.
	The tensor algebra in appendix \ref{app:4pf} was performed using
	the xAct package for Mathematica~\cite{xAct}.

	\appendix

	\section{Derivation of the Wetterich ERG equation}
	\label{app:app1}	
	
	In the second part of the main body of this paper we have addressed the naturalness question
	of $P(X)$ theories. In \S\ref{sec:ERG} we required the exact
	renormalisation group flow equation as a means to compute the
	quantum corrections to the classical Lagrangian
	to all-orders in loops.
	In this appendix we review the derivation of the Wetterich ERG equation.
	We begin with the definition of the infrared regulated generating functional $W_{\kappa}$ defined by
	\ba
\label{eq:Wkappa}
	e^{W_\kappa[J] }=\int \D [\phi] \, e^{-S(\phi) + J \phi- \int \d^4 x\frac{1}{2} \phi \hat R_\kappa \phi} \, .
	\ea	
	Since the only place the regularisation scale, $\kappa$, enters is through $\hat R_{\kappa}$, we have
	\ba
	\partial_{\kappa} W_{\kappa} = -\left\langle \int \d^4 x\frac{1}{2} \phi \,\partial_{\kappa}\hat R_\kappa \, \phi \right\rangle = - \frac{1}{2}
	\int \d^4 x \int \d^4 y \, \partial_{\kappa} \,  R_\kappa(x,y)  \langle \phi(x)  \phi(y) \rangle \, ,
	\label{eq:derivW}
	\ea
	where $R_\kappa(x,y) = \hat R_\kappa(x) \delta^4(x-y)$ and the angle brackets denote the path integral average
	\ba
	\langle O \rangle = \frac{\int \D [\phi] \, O \, e^{-S(\phi) + J \phi- \int \d^4 x\frac{1}{2} \phi \hat R_\kappa \phi} }{\int \D [\phi] \, e^{-S(\phi) + J \phi- \int \d^4 x\frac{1}{2} \phi \hat R_\kappa \phi} } \, .
	\ea
	Since $W_\kappa[J]$ is a generating functional it determines the two-point function
		\ba
	\langle \phi(x)  \phi(y) \rangle = \frac{\delta^2 W_\kappa[J] }{\delta J(x) \delta J(y)} + \langle \phi(x) \rangle \langle \phi(y) \rangle \, .
	\label{eq:2pffunctional}
	\ea
	Defining the Legendre transform of $W_\kappa[J]$ via
	\ba
	\tilde \Gamma_\kappa[\bar \phi] =   -W_\kappa[J]+ J(x) \bar \phi(x)  \,  ,
	\ea
	where $\bar \phi = \langle \phi \rangle$,
	then taking $\bar \phi$ to be independent of $\kappa$ (which implies $J$ is dependent of $\kappa$) and differentiating we have
	\ba
	\partial_\kappa \tilde \Gamma_\kappa[\bar \phi] = -\partial_\kappa W_\kappa[J]|_J - \int \d^4 x \left(- \frac{\delta W_\kappa[J]}{\delta J(x)}+ \bar \phi \right) \partial_\kappa J(x) = -\partial_\kappa W_\kappa[J]|_J \, .
	\ea
	The two-point function $\langle \phi(x)  \phi(y) \rangle$ may also be obtained from $\tilde \Gamma_\kappa$ via
	\ba
	\int \d^4 z \, \frac{\delta^2 \tilde \Gamma_\kappa}{\delta \bar \phi(x) \delta \bar \phi(z)}  \frac{\delta^2 W_\kappa[J] }{\delta J(z) \delta J(y)} = \delta^4(x-y) \, .
	\ea
	In index suppressed notation, from Eq. \eqref{eq:2pffunctional}, we symbolically write
	\ba
	\langle \phi \phi \rangle - \bar \phi \bar \phi=  \frac{\delta^2 W_\kappa[J] }{\delta J^2
	} = \left[\frac{\delta^2 \tilde \Gamma_\kappa}{\delta \bar \phi^2
	}\right]^{-1} \, .
	\ea
	Putting this together into Eq. \eqref{eq:derivW} we obtain the flow equation for $\tilde \Gamma_\kappa$
	\ba
	\partial_{\kappa} \tilde \Gamma_\kappa[\bar \phi] = \frac{1}{2} {\rm Tr} \left[\partial_{\kappa}\hat R_\kappa \frac{1}{\left(\frac{\delta^2 \tilde \Gamma_\kappa}{\delta \bar \phi \delta \bar \phi}\right)}\right]+ \int \d^4 x\,\frac{1}{2} \bar \phi \partial_{\kappa}\hat R_\kappa \bar \phi \, .
	\ea
	Finally for convenience we define the effective averaged action $\Gamma_\kappa$ via
	\ba
	\Gamma_\kappa[\bar \phi] = \tilde \Gamma_\kappa[\bar \phi] - \int \d^4 x\frac{1}{2} \bar \phi \hat R_\kappa \bar \phi \, ,
	\ea
	so that the final form of the ERG equation is (dropping the bar on $\phi$)
	\ba
	\partial_{\kappa}  \Gamma_\kappa[ \phi] = \frac{1}{2} {\rm Tr} \left[\partial_{\kappa}\hat R_\kappa \frac{1}{\left(\frac{\delta^2  \Gamma_\kappa}{\delta  \phi \delta \phi}+ \hat R_\kappa \right)}  \right]\, .
	\ea
	This is the form used in the main text in \S\ref{sec:ERG} for which
	\ba
	e^{-\Gamma_\kappa(\phi)} =\int\D [\chi] \, e^{-S(\phi+\chi) + \frac{\delta \Gamma_k(\phi)}{\delta \phi} \chi -\int \d^4 x \frac{1}{2}
	\chi \hat R_{\kappa} \chi} \,.
	\ea

	\section{Dimensionless couplings analysis}
	\label{app:dimcouplings}

In this appendix we re-derive the quantum stability argument in the large kinetic term regime of \S\ref{sec:LeadingOrderDerivatives}. We will  only assume that the derivative interactions dominate over the standard kinetic term where the $P(X)$ theory is defined at $\Lambda_c$ and make no further assumption at different values of $\kappa$.

We start with the ERG in its dimensionless form derived in Eq.~\eqref{eq:dimensionlessERG3}
\ba
\frac{\partial \tilde P_\kappa(\tilde X_\kappa)}{\partial \ln \kappa}+ 4 \tilde P_{\kappa}(\tilde X_\kappa)- 4 \tilde P'_{\kappa}(\tilde X_\kappa) \tilde X_\kappa &=&  \frac{1}{ (2 \pi)^4} \int_{|q|<1} \d^4 q   \left( \frac{\mathcal{Z}_\kappa}{\mathcal{Z}_\kappa+\(\tilde{Z}_{\kappa}^{\mu \nu}-\mathcal{Z}_\kappa \delta^{\mu\nu}\) q_{\mu} q_{\nu}} \right) \label{eq:dimensionlessERG2} \\
 &=& \frac{1}{ (2 \pi)^4} \int_{|q|<1} \d^4 q   \left( \frac{1}{1+\(\hat{\tilde{Z}}_{\kappa}^{\mu \nu}- \delta^{\mu\nu}\) q_{\mu} q_{\nu}} \right)\,,
\label{eq:dimensionlessERG3}
\ea
where similarly to \S~\ref{sec:LeadingOrderDerivatives}, we define $\hat{\tilde Z}_\kappa ^{\mu\nu} \equiv \tilde{Z}_\kappa ^{\mu\nu}/\mathcal{Z}_\kappa$.
We recall here again that we define our $P(X)$ theory at $\Lambda_c$. This means that $\mathcal{Z}_{\Lambda_c}=1$ and $\hat{\tilde Z}_{\Lambda_c} ^{\mu\nu} \equiv \tilde{Z}_{\Lambda_c} ^{\mu\nu}$.

For simplicity, we focus here on the  case where  $\tilde Z^{\mu\nu}$ is conformal, $\tilde Z_\kappa^{\mu\nu}=\tilde Z_\kappa \delta^{\mu\nu}$, then we find
\ba
\frac{\d }{\d \ln \kappa}\left[\kappa^4 \tilde P_\kappa(\tilde X_\kappa)\right]&=&\kappa^4\left[ \frac{\partial \tilde P_\kappa(\tilde X_\kappa)}{\partial \ln \kappa}+ 4 \tilde P_{\kappa}(\tilde X_\kappa)- 4 \tilde P'_{\kappa}(\tilde X_\kappa) \tilde X_\kappa \right]\\
&=& \frac{\kappa^4}{2^4 \pi^2}\frac{\(\hat{\tilde Z}_\kappa-1\)-\log \hat{\tilde Z}_\kappa }{\(\hat{\tilde Z}_\kappa-1\)^2} \, .
\label{eq:derP}
\ea
This equation can be integrated to give
\ba
\Lambda^4 P_0(X)= \Lambda^4 P_{\Lambda_c}(X) -\frac{1}{2^4 \pi^2 }
\int_{-\infty}^{\ln \Lambda_c} \d \ln \kappa \ \kappa^4  \frac{\(\hat{\tilde Z}_\kappa-1\)-\log \hat{\tilde Z}_\kappa }{\(\hat{\tilde Z}_\kappa-1\)^2} \,.
\ea
Now performing a Taylor expansion of the integrant about $\kappa=\Lambda_c$,
\ba
 \frac{\(\hat{\tilde Z}_\kappa-1\)-\log \hat{\tilde Z}_\kappa }{\(\hat{\tilde Z}_\kappa-1\)^2}  & \approx &
  \frac{\(\tilde Z_{\Lambda_c}-1\)-\log \tilde Z_{\Lambda_c}}{\(\tilde Z_{\Lambda_c}-1\)^2} +
  \mathcal{O}\(\p_\kappa  \hat{\tilde Z}_\kappa|_{\Lambda_c}\) (\kappa- \Lambda_c)+\cdots   \\
  &\approx &  \frac{1}{\tilde Z_{\Lambda_c}} +
  \mathcal{O}\(\frac{\p_\kappa  \hat{\tilde Z}_\kappa|_{\Lambda_c}}{\tilde Z_{\Lambda_c}^{2}}\) (\kappa- \Lambda_c)+\cdots\,,
\ea
where in the second line we have used the assumption that within the large kinetic term regime $|\tilde Z_{\Lambda_c}|\gg 1$, the relation \eqref{eq:Zdimensionless}, and we recall that $\tilde Z_{\Lambda_c} = Z_{\Lambda_c}$.

Putting these relations together we obtain
\ba
\label{eq:res2}
\Lambda^4 P_0(X)= \Lambda^4 P_{\Lambda_c}(X) -
\frac{ \Lambda_c^4}{2^6 \pi^2  Z_{\Lambda_c}}  + \Lambda_c^4  \mathcal{O}\( \frac{\Lambda_c \p_\kappa  \hat{\tilde Z}_\kappa|_{\Lambda_c}}{ Z_{\Lambda_c}^{2}} \)\,.
\ea
Now using the relation \eqref{eq:derP} to estimate $ \p_\kappa  \hat{\tilde Z}_\kappa|_{\Lambda_c} $ based on $ \p_\kappa  \tilde P_\kappa|_{\Lambda_c}$ (we recall the flow in $Z^{\mu\nu}$ is determined by the flow in $P_\kappa$), we infer
\ba
\Lambda_c \p_\kappa  \hat{\tilde Z}_\kappa|_{\Lambda_c}  \approx \frac{1}{X Z_{\Lambda_c}}\left(\frac{\Lambda_c}{\Lambda}\right)^4\,,
\ea
so the terms neglected  in \eqref{eq:res2} are suppressed by two additional powers of $Z_{\Lambda_c}$ which is large well within the strong coupling region and are sufficient to compensate the additional powers of  $\(\Lambda_c/\Lambda\)$. Taking for instance the case of the $P(X)$ screening introduced in  \S~\ref{sec:P(X)screening}, with a strong coupling scale of the dark energy scale, $\Lambda\sim$ meV and a cutoff $\Lambda_c\sim$ eV, then the terms neglected  in \eqref{eq:res2} are indeed suppressed compared to the leading terms for almost the entire solar system.

This means that the order of magnitude of the difference between $P_0(X)$ and $P_{\Lambda_c}(X) $ is suppressed by at least one power of $Z_{\Lambda_c}$ and well estimated by the leading contribution $\Lambda_c^4 / Z_{\Lambda_c}$. The point of this analysis is thus to show that even though the corrections are quartic in the scale $\Lambda_c$, they are suppressed by a negative power of $Z_{\Lambda_c}$ which is large deep within the Vainshtein region. Next-to-leading order corrections also come suppressed by higher powers of $Z$ and therefore do not spoil this result.

\section{A simple toy model}
\label{app:4pf}

The simplicity of the Lagrangian which includes the logarithmic quantum corrections in Eq. \eqref{eq:quantumL} may be unsettling.
To reinforce its validity we could have derived it by performing a perturbative
analysis.
To obtain the individual operators in terms of a sum of
Feynman diagrams and then covariantise the result
would be a herculean task.

So for simplicity, we consider in what follows the first term in such a perturbative approach for a simple toy-model and compare the result with that obtained in \eqref{eq:quantumL}. The model we will
investigate is
\ba
P(X)=\frac12 X+\lambda X^2\,,
\label{eq:toyPX}
\ea
or equivalently,
\ba
\label{eq:Ltoy}
\L=-\frac 12 (\p \phi)^2+\frac{\lambda}{\Lambda^4} (\p \phi)^4\,,
\ea
where $\lambda$ is some positive\footnote{\label{footnote1}Since we only want to focus on the radiative stability of the classical theories, we choose the sign of $\lambda$ appropriately so that it does not generate other possible issues with the theory. To be more precise, the positivity of this coefficient is tied with a well-defined local $S$-matrix~\cite{Dvali:2010jz}.} coupling constant.
We exemplify how quantum operators are generated
by explicitly computing one-loop diagrams in the theory given by the Lagrangian \eqref{eq:toyPX} using dimensional-regularisation.

The lowest $n$-point function which can be corrected by quantum fluctuations to \eqref{eq:toyPX}
is the $2$-point function as depicted in Figure~\ref{fig:2pf}.
The background field is massless  and the amplitude of the one-loop contribution associated to the diagram in Figure~\ref{fig:2pf} is forced to vanish in dimensional regularisation.
\begin{figure}[!htb]\vspace{20pt}
\begin{center}
$\mathcal{A}^{\rm (2pt)}\ \ =$\hspace{15pt}
\begin{fmffile}{2pf}
\parbox{30mm}{\begin{fmfgraph*}(50,50)
                  \fmfsurround{i1,i2}
                   \fmflabel{$\phi$}{i1}
 \fmflabel{$\phi$}{i2}
                \fmf{plain,label=$\mathbf{p}$,label.side=left}{i1,v1}
                \fmf{plain,label=$\mathbf{p}$,label.side=right}{i2,v1}
                \fmfdot{v1}
                \fmf{plain,label=$\mathbf{k}$,right=0.5,tension=0.6}{v1,v1}
\end{fmfgraph*}}
\end{fmffile}
\end{center}
\caption{One-loop contribution to the
2-point function.}
\label{fig:2pf}
\end{figure}\\
Hence the Lagrangian \eqref{eq:toyPX}
does not logarithmically correct the $2$-point function at one-loop. This is a well-known
result that massless fields have a vanishing tadpole.

\begin{figure}[!htb]\vspace{20pt}
\begin{center}
$\mathcal{A}^{\rm (4pt)}\ \ =$\hspace{10pt}
\begin{fmffile}{4ptf}
\parbox{20mm}{\begin{fmfgraph*}(80,70)
	             \fmfleft{i1,i2}
	            \fmfright{i3,i4}
	            \fmflabel{$\phi$}{i1}
	            \fmflabel{$\phi$}{i2}
	            \fmflabel{$\phi$}{i3}
	            \fmflabel{$\phi$}{i4}
                \fmf{plain,label=$\mathbf{p}_1$}{i1,v1}
                \fmf{plain,label=$\mathbf{p}_2$}{i2,v1}
                \fmf{plain,label=$\mathbf{p}_3$,label.side=right}{i3,v2}
                \fmf{plain,label=$\mathbf{p}_4$,label.side=left}{i4,v2}
                \fmfdot{v1,v2}
                 \fmf{plain,left=0.7,tension=0.3,label=$\mathbf{k}$}{v1,v2}
                  \fmf{plain,left=0.7,tension=0.3,label=$\mathbf{q}$}{v2,v1}
\end{fmfgraph*}}
\\[20pt]
\end{fmffile}
\end{center}
\caption{One-loop contributions to the $4$-point function. By
conservation of $4$-momentum, it follows that
$\mathbf{q}$=$\mathbf{k}$-$\mathbf{p}_1$-$\mathbf{p}_2$=$\mathbf{p}_4$+$\mathbf{p}_3$-$\mathbf{k}$.}
\label{fig:4pf}
\end{figure}
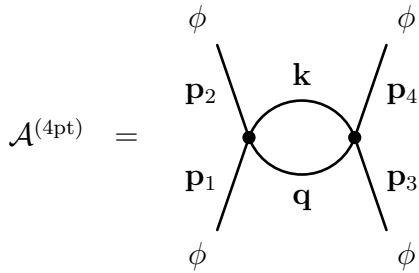

\para{Four-point function}Next we look at the $4$-point function.
The corresponding Feynman diagram is depicted in
Figure~\ref{fig:4pf}.

We label the external legs with different momenta,
$\mathbf{p}_1$, $\mathbf{p}_2$ and $\mathbf{p}_3$, subject to $4$-momentum conservation.
The amplitude associated with
this process is thus
\ba
\mathcal{A}^{\textrm{(4pt)}}\sim& \frac{\lambda^2}{\Lambda^8}\circlearrowleft
\displaystyle{\int{\dfrac{\textrm{d}^4 \mathbf{k}}{(2\pi)^4}}}\dfrac{\left[(\mathbf{p}_1\cdot \mathbf{p}_2)\, (\mathbf{q}\cdot \mathbf{k}) +2 (\mathbf{p}_1\cdot \mathbf{k}) (\mathbf{p}_2\cdot \mathbf{q})\right] \
\left[
(\mathbf{p}_3\cdot \mathbf{p}_4)\, (\mathbf{q}\cdot \mathbf{k}) +2 (\mathbf{p}_3\cdot \mathbf{k}) (\mathbf{p}_4\cdot \mathbf{q})
\right]}{\mathbf{q}^2 \ \mathbf{k}^2} \nn
\ea
where the sum $\circlearrowleft$ is performed over all the cyclic permutations of momenta.
Using dimensional-regularisation, we indeed recover the result from \eqref{eq:quantumL} expanded to the same order,
\begin{equation}
\mathcal{L}_{\rm dim-reg} \sim \frac{\lambda^2}{\Lambda^8} \
(\p\phi)^2 \
\left\{
 \p^{\alpha}\phi \Box^2\p_{\alpha}\phi
+4 \Box \phi \Box^2\phi
+3  \Box\p_{\alpha}\phi \Box\p^{\alpha}\phi
\right\} \,.
\label{eq:correct4pf}
\end{equation}
As expected, we observe the higher derivative terms emerging at the quantum level.

\subsection*{The rising of a ghost?}
The operators generated at one--loop in \eqref{eq:correct4pf}
 are \emph{not} a total derivative, and thus are not redundant
in the technical sense.
The reader might be worried
that the one--loop effective Lagrangian generated quantum mechanically now contains operators which have more than two derivatives
acting on the fields, which would signal the presence of a ghost.
We stress that quantum effects will {\it inevitably} generate
higher derivatives terms (like in GR).
Higher derivatives would be unacceptable if they led to an Ostrogadski instability, or in other words if they produced a new pole in the propagator.

Let us focus, for example, on the operator
$(\p^2 \phi)^4/\Lambda^8$. We can expand it about an arbitrary
background, $\phi_0$, and deduce that the mass of the would be ghost is
$m_{\rm ghost}\simeq \Lambda^4/(|\p^2\phi_0|)$. However, as
we have argued in the main text,
we can design background configurations for which $|\p\phi_0|\sim \Lambda^2$,
provided $|\p^2\phi_0|\ll\Lambda^3$. This condition ensures both the
radiative stability of the theory, as well as the effective absence
of ghosts at energy scales which could be probed by this EFT.

\section{Generalisation to higher-loops}
\label{appendix:moreloops}

In part~\ref{part:validity} of this paper we have quoted the formula for the logarithmic corrections induced by quantum effects.
The result presented in Eq.~\eqref{eq:quantumL} is valid at one-loop. We now generalise this argument to an arbitrary number of loops $\ell$ and focus again on the running of the operator coefficients. It is understood that all the statements below apply to the finite contributions as captured by the logarithms.

For an arbitrary $P(X)$ model, since the field has no mass (nor potential), one can never generate a running of the zero-point function (\ie, cosmological constant) nor of a potential for the scalar field (as is well known, the running of the cosmological constant only comes from massive fields). For a $P(X)$ model, we have seen that all the finite contributions involve higher derivatives of the scalar field.
In what follows we generalise this argument to an arbitrary number of loops.

Consider a generic $P(X)$ model, which can be written as
a series such as $\mathcal{L}=\sum_m\, \lambda_m\, \Lambda^4 X^m$ and let us compute a $(2n)$-point function.  At the very least, to have a finite contribution, this diagram must have $M\ge 2$ vertices of the form $X^{m_j}$ with $j=1,\cdots,M$ and must involve $\ell$-loops, with
\ba
\label{eq:number of loops}
\ell=r-M+1-n\ge 1\,,
\ea
with $r=\sum_{j=1}^M m_j$,
following Euler's formula.
Then on simple dimensional grounds, such a diagram has finite amplitude of the form
\ba
 \mathcal{A}^{(n{\rm-pt})}_{r} \sim \frac{p^{4(1+r-M)-2n}}{\Lambda^{4(r-M)}} \,,\nn
\ea
where $p$ plays the role of the external momentum, which translates into the following operator
\ba
\L_{n, m_j}\sim \frac{\p^{4(1+r-M)-2n}}{\Lambda^{4(r-M)}} \phi^{2n}\,.
\label{eq:Lmoreloops}
\ea
The result in Eq.~\eqref{eq:Lmoreloops} is much more powerful
and reinforces the
results at one-loop. Indeed,
from Eq.~\eqref{eq:number of loops} we have $(1+r-M)= n+\ell$ we immediately infer that the number of derivatives in the $(2n)$ fields is $2n+4\ell$, which inevitably means that there is always more than one derivative per field. We can always express
these operators (symbolically) as $f(X)(\p^{2\ell+1}\phi)^2$. Remarkably, the number of derivatives per
field increases with the number of loops. This means that  in the derivative expansion higher order loops are even more suppressed.

\section{The cubic Galileon: an illustrative example of a higher-order derivative theory}
\label{app:cubicgalileon}
Our analysis in this paper is primarily focused on $P(X)$ theories, where the Lagrangians only depend on the first derivative of the scalar field. However, our results can be readily generalised to Galileon theories. These theories are very rich phenomenologically and their most interesting regime is that of large non-linearities for which screening solutions of fifth-forces exist.

As before, there are a number of ways
of computing the quantum corrections
in these Galileon models, namely using the
point-splitting technique \cite{Brouzakis:2013lla}, or performing canonical normalisation
and substituting into the Coleman--Weinberg effective potential formula
\cite{Luty:2003vm,Nicolis:2004qq}.
On the other hand, the quantum effective action \eqref{eq:quantumL}
allows for a direct derivation of the covariant version of the Galileon non-renormalisation
theorem. This is precisely what we shall do in this appendix.

Consider the cubic Galileon.
This is the simplest of the Galileon operators, and for the purposes of our discussion it suffices to apply the results to this case. Starting with
the Lagrangian \eqref{eq:galileons} we take $c_4=c_5=0$, and it simply reads
\begin{equation}
\mathcal{L}=c_2 (\p\phi)^2-3c_3\Box\phi (\p\phi)^2\ ,
\label{eq:cubicgalileon}
\end{equation}
where the Lorentzian signature was used in the contraction of the Levi--Civita symbols,
and $\Box\equiv \eta^{\mu\nu}\p_\mu\p_\nu$.
We can fix $c_2=-1/2$, so that $\phi$ is canonically normalised,
and assume $c_3<0$ for stability requirements under quantum
corrections to be met (see footnote \ref{footnote1}). Using the
background field method of \S\ref{sec:fieldexp}, we can identify the elements
in the kinetic operator \eqref{eq:defZ}
\begin{equation}
Z^{\mu\nu}[\phi_0]=\delta^{\mu\nu}+12 c_3
\left(
\dfrac{\Box\phi_0}{\Lambda^3} \delta^{\mu\nu} -\dfrac{\p^{\mu}\p^{\nu} \phi_0}{\Lambda^3}
\right)\ ,
\label{eq:Zcubicgalileon}
\end{equation}
where  all the boundary terms have been discarded in the process.

Then the quantum corrections given
in Eq.~\eqref{eq:quantumL} are simply a function of the curvature invariants built out of
the effective metric given by Eq.~\eqref{eq:effmetric}.
In analogy to the conclusions of \S\ref{sec:1PI},
the Ricci curvature tensor involves terms of the schematic form
\begin{equation*}
\textrm{Ricci}\sim 
\dfrac{\textrm{d}^2 Z}{Z}
+ \left(\dfrac{\textrm{d}Z}{Z} \right)^2\ .
\end{equation*}
The formula above agrees with the analysis of Refs. \cite{Luty:2003vm,Nicolis:2004qq},
which cited the quantum corrections as being schematically of the form
\begin{equation}
\mathcal{L}^{\textrm{quantum}}_{\textrm{log}}\supseteq
\left[
\dfrac{\textrm{d}^2 Z}{Z}- \frac{1}{2}\left(\dfrac{\textrm{d}Z}{Z} \right)^2\right]^2\ ,
\label{eq:Zforgal}
\end{equation}
by arguing that $Z^{\mu\nu}\sim Z\delta^{\mu\nu}$.
Notice that from Eq.~\eqref{eq:Zcubicgalileon} the kinetic operator $Z^{\mu\nu}$ for the cubic Galileon involves operators with two
derivatives acting on the fields (the same will be true for the other
Galileon terms in the Lagrangian \eqref{eq:galileons}), while the quantum corrections introduce operators which are at least one higher order in derivative counting.

Therefore, we recover the usual result for Galileons: focusing on the logarithmic divergencies, the EFT defined by the Lagrangian \eqref{eq:cubicgalileon} is well defined provided $\phi\sim \Lambda$, $\partial \phi \sim \Lambda^2$ and $\partial^2\phi \sim \Lambda^3$, while $\partial^{n}\phi \sim \Lambda^{n+1}$. This hierarchy between derivatives of the fields ensures that quantum corrections are kept under control. To be more rigorous, the EFT for the cubic Galileon is defined by the regime for which
\begin{equation}
\sqrt{g_{\rm eff}}\ |R^2[g_{\rm eff}]| \ll
\left|
\dfrac{\partial^2 \phi_0}{\Lambda^3}
\right| \, (\partial\phi_0)^2 \, ,
\label{eq:criteriongal}
\end{equation}
where the RHS is rather symbolic (the complete expression should be read from the RHS of Eq.~\eqref{eq:Zcubicgalileon}.

As noted in \S\ref{sec:1PI}, if we use the power-law divergencies as indicators of high-energy dependence, then the quantum corrections will read symbolically as
\begin{equation}
\Gamma^{\textrm{1-loop}}_{\textrm{power-law}}\sim \frac{1}{(4\pi)^2}
\int{\textrm{d}^4 x \sqrt{g_{\textrm{eff}}}} \
\bigg\{\Lambda_c^4
+
\Lambda_c^2 \ \frac{R}{6}
\bigg\}  \, ,
\label{eq:quantumLpowerlawcubicG}
\end{equation}
where $R$ is the Ricci scalar built out of the effective metric $g_{\mu\nu}^{\textrm{eff}}=\sqrt{g^{\textrm{eff}}} \, Z_{\mu\nu}$ and $Z_{\mu\nu}$ is the inverse of the kinetic operator $Z^{\mu\nu}$ in Eq.~\eqref{eq:Zcubicgalileon}. As soon as we consider solutions inside the Vainshtein radius, the corrections generated by power-law divergencies excite operators of the same form as the Galileon ones originally present in the classical action. In part~\ref{part:validity} of this paper we discarded this family of divergencies, for the reasons explained in \S\ref{sec:1PI}. Applying the same arguments to the Galileons, the quantum corrections
in Eq.~\eqref{eq:quantumLpowerlawcubicG} can be dismissed as not providing an accurate accounting of high-energy physics effects.
\vspace*{0.3cm}

\section{A closer look at DBI: a symmetry manifest approach}
\label{appendix:5d}
In the main text we have discussed the features of DBI as a four-dimensional EFT. In fact, DBI arises in the context of higher-dimensional brane models, as a nontrivial combination of the Dirac and the Born--Infeld actions, where the reparametrisation invariance
is made manifest. In this appendix we investigate whether performing the calculations of the quantum corrections in a higher dimensional setup offers special (if any) insights.

\subsection{Where did the symmetry go?}
\label{subsec:sym}

All the terms in the one-loop effective action \eqref{eq:quantumL} trivially satisfy the shift symmetry of $P(X)$ Lagrangians. One could wonder if
other symmetries in the classical action are also preserved at the level of the quantum effective action in \eqref{eq:qea}.

To address this question, we consider the special
example of DBI, as briefly introduced in \S\ref{sec:introduction}.
With a higher-dimensional motivational setup, the DBI action describes the
relativistic motion of a brane moving in a generically
warped geometry.
We suppose for simplicity the brane moves along
a cut-off throat, to mimic the absence of warping.
The DBI Lagrangian in this case is given by
\begin{equation}
S_{\rm DBI}=\int{\textrm{d}^4 x} \
\left\{
-\Lambda^4 \sqrt{1-X}+\Lambda^4
\right\} \,,
\label{eq:dbiactionA}
\end{equation}
where again $X=-(\partial\phi)^2/\Lambda^4$.

Not only is this theory invariant under the shift symmetry,
but it is also invariant under a non-linear diffeomorphism given in Eq.~\eqref{eq:dbisym}.
In fact, DBI is the only model within the class of $P(X)$
theories which is invariant under this non-linear symmetry.
For small $X$ the Lagrangian \eqref{eq:dbiactionA}
reproduces the theory of a canonically normalised scalar
field, with the first interaction being
of the form modelled in Eq.~\eqref{eq:toyPX}. But the most interesting
regime is that of large self-interactions
measured by powers of $X$. The
presence of the square root in \eqref{eq:dbiactionA} provides a means to resum
an infinite tower of such interaction channels within the strong coupling regime of the theory. In that case and following the
terminology in Eq.~\eqref{eq:defirrelev}, we can say that DBI
contains an infinite number of irrelevant but important
operators, of the form $X^n$, where $n$ runs from $1$ to infinity.

Does the quantum effective action \eqref{eq:quantumL}
satisfy the DBI symmetry whose infinitesimal form is
\eqref{eq:dbisym}? Explicit verification shows that it does not,
which might be indication of trouble and hint at a lack of consistency of our result.
Indeed, one expects that invariance of the classical action
under a certain symmetry should
be respected by quantum effects and therefore be manifest at
the level of the quantum effective action.
For most cases, both the Lagrangian density as well as the
measure of the path integral remain in fact invariant under the symmetry
transformation.
Nevertheless, exceptions exist and, in particular, when the
symmetries are non-linearly realised, the invariance under the
symmetry is not preserved at the quantum level ~\cite{Weinberg:1996kr}.

One way of understanding how to preserve the symmetry
\eqref{eq:dbisym} under quantum corrections is to
notice that the formula for $Z^{\mu\nu}$ in
Eq.~\eqref{eq:defZ} can have an origin in higher-dimensional models.
Indeed, it is conformally related to the metric induced on a
probe brane immersed in a higher-dimensional space-time~\cite{Silverstein:2003hf,deRham:2010eu}
\begin{equation}
Z_{\mu\nu}=\Omega^2(X) \, q\mn \hspace{20pt}{\rm with}\hspace{20pt} q\mn=
\delta_{\mu\nu}+\frac{1}{\Lambda^4}\p_{\mu}\phi_0 \p_{\nu}\phi_0\, .
\label{eq:disformal/induced}
\end{equation}
The induced metric $q\mn$ appropriately transforms as a tensor
under the DBI symmetry
associated with boosts and rotations in the extra dimension,
as described by the non-linear transformation
\eqref{eq:dbisym}.

If $Z_{\mu\nu}=q_{\mu\nu}$, or equivalently $\Omega^2=1$,
then $Z_{\mu\nu}$
and scalar quantities constructed from it
would be explicitly invariant under the transformation
in Eq.~\eqref{eq:dbisym}.
However, because of the $X$-dependence of the conformal
factor $\Omega$, $Z_{\mu\nu}$ and therefore the effective
metric do not transform as tensors under the transformation.
The degree of breaking of the symmetry will be measured by
operators originated from terms such as
$(\p\Omega/\Omega)^2 \sim (\p Z/Z)^2$,
and similar derivatives as we have deduced in Eq.~\eqref{eq:Zloop},
at the level of the quantum corrections.

Ultimately to keep a prescription where the symmetry is made manifest one should rather work in the higher-dimensional setup, where the DBI symmetry originated from.

\subsection{DBI from a five-dimensional embedding}
\label{subsec:hdim}

In what follows we consider a probe-brane located at $x^5=\phi(x^\mu)$ in the flat-slicing of five-dimensional Minkowski (or Euclidean space). The induced metric on the brane is thus given by
\begin{equation}
q\mn = \delta\mn-\frac{1}{\Lambda^4}\p_\mu \phi \p_\nu \phi\,.
\label{eq:inducedmetric}
\end{equation}
The inverse of the induced metric on the brane is simply given by
\begin{equation}
q^{\mu\nu} = \delta^{\mu\nu}+\frac{\gamma^2}{\Lambda^4}\p^\mu \phi \p^\nu \phi\,,
\end{equation}
where indices are raised and lowered using $\delta^{\mu\nu}$ and
with
\ba
\gamma\equiv\frac{1}{\sqrt{1-X}}\ ,
\label{eq:defgamma}
\ea
being the Lorentz boost factor.

In five dimensional GR with a brane, there will be bulk loops and brane loops. Performing again a one-loop effective action, one can check that the bulk loops take the form
\ba
\label{1loop bulk}
\mathcal{L}_{\rm QC\ bulk}^{(5d)}\sim \sqrt{G}\, M_5^5\(R_{ABCD}^{(5)}[G]\)^2 \,,
\ea
where $M_5$ if the five-dimensional Planck scale,  $G_{AB}$ is the five-dimensional metric
with $5$-dimensional indices $A,B,\cdots$,
 and $(R^{(5)}_{ABCD})^2$ designates five-dimensional contractions of the  Riemann tensor.
At higher orders in loops we generate even higher derivative terms of the form
\ba
\label{1loop branefirst}
\mathcal{L}_{\rm QC\ bulk}^{(5d)}\sim \sqrt{G}\, M_5^5   \sum_{ n, m \ge 0} \left( \frac{\nabla}{M_5} \right)^n  \left( \frac{{}^{(5)}\!R_{ABCD}}{M_5^2} \right)^{m}\,.
\ea
Additionally, the finite part of the brane loops take the form
\ba
\label{1loop brane}
\mathcal{L}_{\rm QC\ brane}^{(4d)}\sim \sqrt{q} \sum_{\ell, n, m \ge 0} \Lambda^{4-2\ell -2n -m} \, R^\ell \, \grad^{2n} \, K^{m} \,,
\ea
where $R$ and $\grad$ are derived with respect to the induced metric $q\mn$, which has determinant denoted by $q$.
In the limit $M_5\to \infty$ keeping $\Lambda$ finite, the bulk loops completely decouple while the brane loops remain.
Here $K\mupn$ represents the extrinsic curvature given by~\cite{deRham:2010eu}
\ba
K\mupn=-\frac{1}{\Lambda^2}\, q^{\mu\alpha}
\, \gamma \, \bar{\grad}_\alpha \bar{\grad}_\nu \phi\,.
\ea
where $\bar{\nabla}$ is to be understood as the
covariant derivative with respect to the metric $\delta_{\mu\nu}$.
For cartesian coordinates, this is simply the
usual partial derivative, but whenever the coordinate system is not
cartesian, there will be important differences.

Notice that in this formalism both the bulk and the brane loops are manifestly invariant under the DBI symmetry. Indeed, the induced metric, the extrinsic curvature and the five-dimensional Riemann tensors all transform as tensors under \eqref{eq:dbisym} and the brane and bulk actions constructed out of scalar quantities are thus manifestly invariant.

\para{Regime of validity of the EFT}Classical solutions computed using the DBI action \eqref{eq:dbiactionA} are within the regime of validity of the theory as long as the contributions from \eqref{1loop brane} are small compared to the operators in \eqref{eq:dbiactionA}.

Power-law divergences include contributions in \eqref{1loop brane} with $\{\ell, n ,m\}=\{0,0,0\}$ corresponding to the equivalent of the cosmological constant problem. If that power-law divergence were taken seriously, DBI would not be technically natural unless the strong coupling scale was identified with the cut-off which is at least $M_5$. If that were true,
the bulk loops would not decouple. In what follows we take the approach that power-law divergences are regularisation and field-dependent  and may not capture the UV physics (see also Ref.~\cite{Burgess:1992gx}). Moreover, we put them under the same category as the cosmological constant problem
until Part~\ref{part:Naturalness} of the paper where naturalness questions are addressed precisely.

Therefore focusing on logarithmic divergences, given by $\{\ell, n ,m\}=\{0,0,4\}$, and regardless of the classical configuration,
all the eigenvalues of  $K\mupn$ should be small compared to the scale $\Lambda$
\ba
\label{validity}
\left|\lambda_K\right| \ll \Lambda\,.
\ea
The most interesting regime of DBI is that of large self-interactions where $|X|\sim 1$ and more specifically when $|X|\to 1$ and  $\gamma \gg 1$, with $\gamma$ defined in Eq.~\eqref{eq:defgamma}.
In that case the criterion \eqref{validity} inferred from the previous symmetry-preserving argument  implies
\ba
\label{eq:CritDBIK}
| \p^2 \phi | \ll \gamma^{-3} \Lambda^3\,,
 \label{eq:highdimcrit}
\ea
where care should be taken in evaluating the double derivative  if the coordinates are not cartesian.

To compare this with the result \eqref{eq:criteriaZ}, which was
derived following a master formula due to Barvinksy \& Vilkovisky,
we start by writing $Z^{\mu\nu}$ given in \eqref{eq:defZ} as
\ba
Z^{\mu\nu}=\gamma \delta^{\mu\nu}-\gamma^3\,\frac{ \p^\mu \phi_0\p^\nu \phi_0 }{\Lambda^4}\,.
\ea
In the regime where $\gamma \gg 1$, the smallest eigenvalue of $Z$ goes as $\lambda_{\rm min}\sim \gamma$, while the largest goes as $\lambda_{\rm max}\sim \gamma^3$. Using the criterion \eqref{eq:criteriaZ} derived from the four-dimensional one (and $\ell$)--loop effective action, we
can in principle infer
how such condition translates explicitly
in terms of the eigenvalues of
$Z^{\mu\nu}$, including when there is a hierarchy between them. The contractions implied in the
expression for the Ricci scalar in Eq.~\eqref{eq:quantumL}
show that the hierarchy of
the eigenvalues only enters in a very peculiar way.
A direct calculation shows that,
at worst, the
eigenvalues $\lambda_{\rm min}$ and $\lambda_{\rm max}$
need to satisfy
\ba
\label{eq:CritDBI}
\Lambda^4 \, \gamma^{-1} \gg
\left(\dfrac{\lambda_{\rm max}}{\lambda_{\rm min}}\right)^{3/2}
 \left|
\frac{\p \lambda_{\rm min}}{\lambda_{\rm min}}
\right|^4
 \sim
 \gamma^3 \left(\frac{\p\gamma}{\gamma}\right)^4
 \, ,
 \label{eq:crit2}
\ea
where the right-hand side is symbolic.
When $|\p\phi| \sim \Lambda^2$, this implies
\begin{equation}
|\p^2 \phi| \ll \gamma^{-3} \Lambda^3\ ,
\label{eq:critgamma}
\end{equation}
which is precisely the same criterion as \eqref{eq:highdimcrit}
found using the five-dimensional embedding picture.

Finally, notice that
in principle the generic criterion  \eqref{eq:criteriaZ} could have  been too restrictive for DBI as it might have included contributions which would not have been generated had one followed a fully higher-dimensional description.
The four-dimensional and the higher-dimensional theories have
different fundamental degrees of freedom, so it is not
surprising the respective quantum corrections might differ. However, on a practical level, if we only keep track of logarithmic divergencies (as was
done in part \ref{part:validity} of this paper), we have shown that the different perspective does not affect our results.

	\bibliographystyle{JHEPmodplain}
	\bibliography{references}

\end{document}